\documentclass{article}
\usepackage[utf8]{inputenc}
\usepackage[a4paper, margin=1in]{geometry}
\usepackage{url}
\usepackage{natbib}
\usepackage{graphicx}
\usepackage{epstopdf, epsfig}
\usepackage{float}
\usepackage{graphicx}
\usepackage{epstopdf, epsfig}
\usepackage{amsmath}
\usepackage{soul}
\usepackage{color}
\usepackage{siunitx}
\usepackage{subcaption}
\usepackage{commath}
\usepackage[dvipsnames]{xcolor}
\usepackage{xfrac}
\usepackage{cancel}
\usepackage{hyperref}
\usepackage{parskip}
\usepackage{csquotes}

\newcommand{\co}{CO$_2$ }

\newcommand{\nl}{$\sum{n}$}

\newcommand{\rnac}{RNA\,copies}

\title{The ventilation of buildings and other mitigating measures for COVID-19: a focus on winter 2020}
\author{The Royal Society `Rapid Assistance for Modelling the Pandemic (RAMP)' project\\Task 7: Environmental and aerosol transmission}
\date{\today}

\begin{document}

\maketitle

\textbf{This document constitutes draft guidance which has been published for consultation purposes only. \\
The intended audience include: advisors to UK Government  (e.g. SAGE), Public Health England, relevant government departments (e.g. the Department for Education), ventilation practitioners (e.g. manufacturers and designers), skilled building service managers, and interested scientists.}

\section{Executive summary} \label{sec:sum}

Winter 2020 presents significant risks in managing the ventilation of buildings and maintaining a healthy indoor environment. The extent of the COVID-19 pandemic, both in terms of its stage of development and the controlling measures in-place, varies widely across the globe both inter- and intra-country. Within the United Kingdom of Great Britain and Northern Ireland (UK) the disease seems to be at a dangerous juncture with the reproductive number (i.e. the average number of infections arising from a single infectious case) currently above one for COVID-19 \citep[e.g. the R-number is being reported, as of September 18, as 1.1--1.4,][]{UKGovR}. This is at a time when cooler weather is approaching where people typically spend longer indoors in the company of others and the supply of outdoor ventilation air is reduced. Simultaneously, the UK and other governments are trying to re-open their economies, which is accompanied by increased social interactions (e.g. schools reopened in August and September 2020 throughout the UK, with universities soon to follow). This report is intended to review much of the knowledge surrounding the indoor spread of COVID-19, and present new results which can inform guidance for mitigating the impact of the disease --- our focus has been the UK but our findings will be more widely useful.
 
It is the premise of this report that COVID-19 may be spread via three main routes (droplet, contact and airborne) all of which we assume to be potentially significant. By consideration of the indoor environment and our behaviour within it, we discuss potential mitigating strategies for all three routes. However, our primary focus is the airborne/aerosol route, since mitigating the spread via this route is the most challenging. The evidence suggests that adequate supply of outside (or at least uncontaminated) air is crucial in helping ensure the reproductive number of a particular indoor space is minimised and ideally remains below one. Findings suggest that for many indoor spaces which are regularly attended by the same/similar group of people (e.g. an open plan office or school classroom) adhering to existing design guidance for adequate ventilation (e.g. 10\,litres per second per person, or 10\,l/s/p, for offices \citep{CIBSEA} and 5--10\,l/s/p for school classrooms \citep[see][]{BB101}) and occupancy densities, results in airborne infection risk that remains relatively low for most reasonable scenarios. Modelling estimates suggest that in these cases the reproductive number is less than one suggesting each infection is expected to give rise to less than one new infection. However, for any indoor space where respiratory activity levels of a potential infector are expected to increase to anything above those corresponding to sedentary talking, increased caution is required.

Appropriate social distancing (\S\ref{sec:soc_dist}) and/or the wearing of face coverings appear to be effective measures to help mitigate the risk of transmission via the droplet route. Increased hand hygiene and the cleaning of surfaces, particularly high-touch public surfaces (i.e., those frequently touched by more than one individual), will reduce transmission via the contact route. Cleaning of surfaces using disinfectants based on alcohols and reduced contact with these surfaces appear to be effective measures to reduce transmission  (\S\ref{sec:contact}). Indoor spaces that bring individuals together over long periods, e.g. open-plan offices, school classrooms and the like, or those that lead to increased respiratory activity, e.g. gyms, choral halls, etc..., are expected to make 
 the airborne spread of the virus an important consideration. 

Assessment of the ventilation provision, or where practical the monitoring of \co levels to indicate ventilation provision (\S\ref{sec:airb}), can help manage the risk of COVID-19 transmission via the airborne route in winter 2020. Most documented cases of transmission which are believed to have arisen from the airborne route have been in environments where the outdoor air supply would not have complied with current UK design guidance. It is inferred from this, and the documented modelling, that provision of outdoor air in-line with existing design guidance will help reduce the risk of transmission by the airborne route. The rate of provision of outdoor air can be inferred by monitoring \co levels, maintaining these below about 1\,000\,ppm being indicative of adequate ventilation in many indoor spaces, including offices \citep[with design guidance for some indoor spaces permitting 1\,500\,ppm see][and \S\ref{sec:VentGuide} for a fuller discussion]{BB101}. However, higher ventilation rates may be needed wherever activity levels increase beyond desk-based work. Risks cans also be reduced by reducing occupancy (whilst ensuring full outdoor air supply rates are maintained), staggering occupancy (via appropriate timetabling) and by the purging of indoor spaces between events (\S\ref{sec:purge}). Numerous additional engineering strategies are available to help further reduce the risk of transmission and we review these in detail (see \S\ref{sec:other}).

The review of current knowledge has identified the following key research questions.
\begin{itemize}

    \item What are the SARS-COV-2 viral load distributions and respiratory droplet size distributions emitted by an individual carrying out activities such as sitting, walking, talking, singing, sneezing, coughing? How might these vary with an individual's size, age, etc...?  
    
    
    
    \item What is the trajectory of droplets and aerosols containing viral particles from exhalation to removal under different ventilation modes, occupancy levels, occupant behaviour/movement, and environmental conditions (e.g. temperature, humidity, etc...)?
    
    \item How can risk and severity of infection of an individual be determined from the nature of viral exposure (i.e. how and where droplets or viral particles are deposited in the respiratory system of an individual, their frequency, the peak/cumulative dose, etc...)? 
    
    \item What processes determine the timescales for purging a space or determining the frequency of cleaning and which environmental condition affect these?
    
    \item What are the quantitative impacts of wearing face coverings?
    
    \item How effective are localised outdoor air supplies and/or purification methods, and what factors affect their results?
    \item How can existing knowledge, and perhaps answers to the above questions, be deployed to better understand and predict the spread of COVID-19, for which high-spreading statistical outlier events (so called `superspreaders') appear to be significant? 

\end{itemize}


\newpage

\tableofcontents

\clearpage

\section{The indoor spread of COVID-19 and winter 2020}

Florence Nightingale is credited with having promoted the idea that the indoor environment plays a critical role in determining health outcomes. Her pioneering work on hospital ward design is still highly relevant with the guiding principles of high ceilings, adequate natural lighting, and sufficient ventilation proving sound design for any indoor environment. However, modern architectural approaches, which often rely on 
mechanical means to condition the environment, may not follow these principles. In this document we focus on two indoor spaces, namely, open-plan offices and school classrooms,  because these constitute spaces which are recurrently attended by the same group of people, are occupied for significant portions of each weekday, are frequently populated at moderate to high occupancy density and do not often adhere to Nightingale's design principles. In short, these settings contain very common spaces in which a significant proportion of the population may potentially be at moderate to high risk of exposure to the COVID-19 corona virus. However, the findings and guidance reported within this document apply generically to almost all indoor spaces and we urge that the guiding principles be widely applied.


\subsection{COVID-19 transmission indoors}

The novel coronavirus disease (COVID-19), which causes respiratory and other symptoms, was declared a pandemic by the World Health Organization (WHO) in March 2020. Transmission of such respiratory infections occurs via virions encapsulated in particles of respiratory secretions (in this case the virus SARS-CoV-2) formed in the respiratory tract of an infected person and spread to other humans via three routes: the droplet route, the contact route and the airborne route \citep{mittal2020flow}. 

The droplet route involves the transfer of respiratory droplets from an infected person to the mucous membranes of a subject, i.e. respiratory droplets land in the mouth, nose or eyes of others. The contact route takes place when respiratory droplets are deposited onto surfaces that are then touched by other people who go on to touch their mouth, nose or eyes before washing their hands. The airborne route (also called aerosol transmission) occurs when exhaled respiratory droplets are small enough to remain suspended in air such that they can be inhaled into the respiratory system of other people.  

At the beginning of the COVID-19 pandemic, a lack of direct empirical evidence on airborne transmission of SARS-CoV-2 highly influenced health policy decisions which were intended to control the pandemic and the public response to it. However, an increasing body of evidence (particularly in poorly ventilated indoor environments), a better understanding of the disease progression, and information on the asymptomatic and pre-symptomatic transmission of the virus strongly support the case for airborne transmission of SARS-CoV-2 virus \citep[see][for discussion and the references therein]{Morawska20,Morawska2020Time}.

In an indoor environment the ventilating flow modulates the transport and advection of any aerosols (including bio-aerosols), pollutants, and \co produced by indoor-sources/occupants and further determines their subsequent removal from within the indoor environment. Traditionally, building ventilation has been studied in the context of thermal comfort and in the last few decades energy efficiency. However, there has been a timely shift of focus and, in addition to energy efficiency and thermal comfort, indoor air quality (and implicitly the removal of any indoor airborne pollutants produced by the occupants) has become a core focus \citep{sloan2020prioritising}. Within our guidance we exploit this cutting-edge knowledge to offer advice to mitigate COVID-19 spread via the airborne route. However, it is essential not to do so at the expense of considering droplet transmission and contact transmission, and measures to minimise the risk of transmission via these routes must be given as much priority as consideration of the airborne route. As such, within this section we review transmission via the droplet and aerosols route (\S\ref{sec:dropaero}) and contact routes (\S\ref{sec:contact}), we then consider the role that social distancing (\S\ref{sec:soc_dist}), face coverings (\S\ref{sec:masks}), and occupancy behaviour (\S\ref{sec:people}) can play in affecting the various modes of transmission. In \S\ref{sec:airb} we address the role of ventilation in influencing the airborne transmission route. We go on to discuss the suitability of other measures in  mitigating the spread (\S\ref{sec:other}) and we present three appendices covering: factors affecting, and modelling considerations for, surface transmission (appendix \ref{app:surfaces}), details of the potential for ultraviolet germicidal irradiation  (UVGI) air disinfection to mitigate COVID-19 transmission (appendix \ref{app:uvgi}), and details of current governmental guidance for face coverings (appendix \ref{app:masks}).

\subsubsection{Transmission via droplets and aerosols} \label{sec:dropaero}

Respiratory diseases are transmitted by exposure to pathogen-laden droplets produced by expiratory events such as breathing, coughing, sneezing, speaking, singing and laughing \citep{stelzer2009exhalation, yan2018infectious}. The expiratory droplets range between {$0.01$}{$-\SI{1000}{\micro\meter}$} \citep{bake2019exhaled}, and conventionally they are classified in two categories; droplets smaller than $\SI{5}{\micro\meter}$ are referred to as {droplet nuclei or aerosols}, whereas droplets larger than $5-\SI{10}{\micro\meter}$ in diameter are classified as respiratory droplets \citep{world2014infection,milton_rosetta_2020}. This somewhat arbitrary size classification implicitly refers to the transmission modes/mechanisms, namely droplet, and airborne transmission. However, the distinction between droplet transmission and airborne transmission determined by a simple cut-off in droplet size neglects a multitude of physical processes crucial to the droplet evolution within an indoor environment. For example, droplets that  are larger than a selected cut-off size at the source may shrink due to evaporation, becoming sufficiently small (before any impact occurs) that they then contribute to airborne transmission. The distinction between droplet transmission and airborne transmission is better explained by the route of infection. Droplet transmission occurs when a subject is exposed to large ($>5-\SI{10}{\micro\meter}$) pathogen-laden droplets expelled by an infected person that come into contact with their mucous membranes. Droplet transmission usually occurs in close proximity (see \S\ref{sec:soc_dist}). While droplets may fall quickly onto a surface close to the source, aerosols are expected to remain airborne for longer periods and can be advected away from the source with ventilation flows leading to what we term `the airborne transmission route'. Therefore, an important aspect of understanding droplet and airborne transmission is the size distribution of the expiratory droplets containing the virus also include water, salts and organic material \citep{kumar2019could}. Droplets and aerosols produced by violent expiratory events such as coughing and sneezing have been investigated and reviewed by several authors, including \citet{yang2007size, bourouiba2014violent}, \citet{bourouiba2020turbulent} and \citet{mittal2020flow}. However, under normal circumstances, the cumulative amount of expiratory fluid and consequently the droplets and aerosols produced by low-frequency intermittent events such as coughing and sneezing are less than that of high-frequency events such as breathing and talking \citep{gupta2010characterizing}.

The studies conducted on disease progression suggest that infectivity of COVID-19 peaks before the onset of symptoms and consequently, preventing pre-symptomatic and asymptomatic transmission is key to containing the spread of the disease \citep{matheson2020does}. At the early stage of SARS-CoV-2 infection, upper respiratory tract symptoms and the presence of high concentrations of SARS-CoV-2 virus in oral fluids are common \citep{wolfel2020virological} which support the recent findings identifying speech droplets to be a potential cause of transmission \citep{stelzer2009exhalation,anfinrud2020visualizing,stadnytskyi2020airborne}.

Conversational speech produces a wide range of droplet sizes (sub-micron up to the order of $\SI{100}{\micro\meter}$
) which are exhaled at speeds of  of the order of $3.5-4\SI{}{\meter\per\second}$. 
The reported size distributions of speech droplets show a large variation, due to different measurement techniques, evaporation of droplets prior to measurement, and natural variation amongst different people \citep{xie2009exhaled}. Aerosol measurements 
capable of measuring particles in the range $\SI{0.5}-\SI{20}{\micro\meter}$ indicate that speech droplets 
form across the measurement range, with geometric mean diameter of $\sim\SI{1}{\micro\meter}$, droplet number concentrations in exhaled breath of the order of $\SI{0.1}-\SI{1}{\centi\meter}^{-3}$, 
and exhaled particle emissions rates of the order or $\SI{1}-\SI{10}{\second}^{-1}$ 
\citep{asadi2019aerosol,johnson_modality_2011}. \citet{asadi2019aerosol} showed that speaking louder is correlated with higher particle emissions and found that a small fraction of people are `super-emitters', who consistently release an order of magnitude more particles than others. \citet{Chao2009} measured the droplet size distribution of cough and speech droplets at mouth opening and found the geometric mean diameter of cough droplets was $\SI{13.5}{\micro\meter}$. In contrast, speech droplets were $\SI{16}{\micro\meter}$, but had a reported maximum diameter of up to $\SI{1000}{\micro\meter}$. \citet{xie2009exhaled} reported the average speech droplet diameter to be between $50-100 \SI{}{\micro\meter}$. Interestingly, this study also showed that both the number and the droplet size increased significantly when the subjects swallowed food dye solution (with or without sugar) before the experiment, indicating that eating may promote the release of higher numbers and larger sizes of expiratory droplets. Although light scattering measures only larger droplets and consequently provides a conservative estimate of total droplet count, \citet{stadnytskyi2020airborne} measured high droplet release rates relative to other studies when using this technique. Both \citet{anfinrud2020visualizing} and \citet{stadnytskyi2020airborne} showed that speech droplets of size $10-100  \SI{}{\micro\meter}$ can remain suspended for up to $\SI{30}{\second}$. Therefore, it is imperative to appreciate that speech droplets can potentially transmit respiratory diseases by both the droplet and airborne transmission routes. In contrast, it has consistently been shown that the majority of aerosol particles in exhaled breath are $< \SI{5}{\micro\meter}$ \citep{fennelly_particle_2020}, which diminishes the possibility of droplet transmission when breathing; however, airborne transmission cannot be ruled out.   

When droplets are exhaled, they evaporate at a rate that depends on droplet size and composition, and the relative humidity and temperature of the air. \citet{redrow2011modeling} compared the evaporation time and resulting nuclei sizes of model sputum, saline solution, and water droplets. They showed that sputum droplets containing protein, lipid, carbohydrate, salt and water leave larger nuclei than salt solution. They calculated the time scales of evaporation of water droplets at room temperature, for  relative humidities between 0 to 80\%, to be $0.1 - \SI{1}{\second}$ for droplets less than  $\SI{10}{\micro\meter}$  and $7 - \SI{40}{\second}$ for $\SI{100}{\micro\meter}$ droplets. Therefore, {it is expected that} droplets larger than $\SI{100}{\micro\meter}$ settle on the floor or other nearby surfaces \citep{liu2017evaporation}, while droplets smaller than about $\SI{10}{\micro\meter}$ tend to form nuclei and are transported as passive scalars, i.e. they were transported by airflows without the dynamics of the airflow being significantly affected \citep{Xie2007}. 

{The final size of exhalation droplets depends upon many factors including the initial size, non-volatile content, relative humidity, temperature, ventilation flow, and the residence time of the droplet. \citet{marr2019mechanistic} gave the equilibrium size for $\SI{10}{\micro\meter}$ sized model respiratory droplets containing $\SI{9}{\milli\gram\per\milli\litre}$ NaCl, $\SI{3}{\milli\gram\per\milli\litre}$ protein, and $\SI{0.5}{\milli\gram\per\milli\litre}$ surfactant to be $\SI{2.8}{\micro\meter}$ and $\SI{1.9}{\micro\meter}$ at relative humidities of $90\%$ and $<64\%$, respectively.}

A significant uncertainty in our ability to quantify the relative importance of airborne transmission is the viral load associated with different aerosol sizes for different expiratory events, at different stages of infection and the potential for natural variation amongst people. This information is currently unknown, leading to large uncertainty bounds.

\subsubsection{Transmission via surface contacts and fomites} \label{sec:contact}

Fomites are inanimate objects that have become carriers of virus particles and these have been shown to play a role in the spread of viruses. Little is known of the true risk of becoming infected by SARS-CoV-2 through this pathway, as it is difficult to isolate from the droplet and airborne routes, the dose required to become infected has not been determined and the majority of studies that have looked at its survival on surfaces have used far greater viral loads than would be deposited naturally \citep{Vasickova2010}. Knowledge of the virus survival on surfaces and interactions between the fomite and droplet/airborne routes through deposition and re-suspension of viral particles is nevertheless important in minimising the risk of infection.

Fomite transmission occurs predominantly from \textbf{human behaviour} through non-infected individuals making contact with or handling infected objects, which may be infected via deposition of large droplets from infected individuals or via direct contact by individuals with viral particles on their hands. The non-infected individual then transfers viral particles from their hands to mucous membranes by touching their eyes, nose or mouth. Advice given by various groups to avoid unnecessary contact between hands and objects in public environments, as well as advice to avoid touching one's face, is sound and should be encouraged. In China it was found that the majority of surfaces within hospital wards that had  infected individuals, were found to have traces of the virus \citep{Ye2020}, and therefore cleaning of surfaces is an important mitigation strategy to avoid infection.

This section briefly summarises what is known about SARS-CoV-2 and its survival and transmission via surfaces. Note that a more thorough and detailed literature review is found in Appendix \ref{app:surfaces} of this document, which details the sources upon which this advice is based.

\mbox{}\par
\noindent \textbf{Factors affecting the survival of SARS-CoV-2 on surfaces}. A number of environmental factors affect the survival of SARS-CoV-2 on surfaces.

\begin{itemize}
\item \textbf{Temperature} effects have been reported to be significant, with higher temperatures decreasing survival times \citep{Dietz2020}. Comfortable indoor temperatures should be maintained and the use of air conditioning should be minimised wherever practical with the appropriate supply of outdoor air remaining a priority. 

\item \textbf{Humidity} has been shown to also have an effect on the virus, with drier conditions being more suitable for virus survival \citep{Biryukov2020}. While higher humidity is preferable to reduce viral infection, there are numerous health issues related to high humidity and promotion of mould growth. We therefore advise that in cold weather the relative humidity should be maintained at between 40--50 \%, rather than below 30 \%, which is typical of many indoor environments in winter \citep{Dietz2020}.

\item \textbf{Light} is also demonstrated as a method for virus deactivation, with UV-C light being shown to deactivate other strains of coronavirus \cite{Bedell2016}. While the use of artificial light cleaning technologies is not suggested as an effective replacement for disinfectant cleaning practices, well-lit rooms, particularly via natural lighting is preferred based on evidence of other viruses \citep{Dietz2020, schuit_airborne_2020}.

\item It is now well-known that SARS-CoV-2 has different survival times on different {surfaces}, with laboratory inoculations of SARS-CoV-2 survival rates varying from 3 hours for paper and tissue to up to 72 hours (3 days) on hard, smooth surfaces such as plastic and stainless steel (and also on surgical masks) \citep{VanDoremalen2020}. Glass and bank notes have survival times in the region of 3 days, with cloth and wood reported at 2 days. While likely viral loads on contaminated objects are not known, and hence risk associated with the fomite pathway relative to that for droplet and aerosol transmission cannot currently be quantified, we believe it is important to regularly clean often-handled objects and surfaces in public spaces. 
\end{itemize}

\noindent \textbf{Cleaning recommendations}. The above leads to the recommendation that it is important to frequently clean door handles, classroom  and meeting room desks, tap handles, swing door handles, ticket machines, pin code keypads, communal office kitchens, etc. Providing point-of-contact public alcohol-based disinfectant, as is now common on university campuses, shops and public transport is an effective mitigation strategy to ensure that public spaces are less likely to become contaminated. We conjecture that an effective mitigation strategy for certain public spaces that involve fomites, such as public computer laboratories found in universities or libraries, would be to encourage users to clean the workspace (keyboards, mouse, desk and hands) before and after use. Further research is needed to establish the efficacy of such interventions depending on the frequency of cleaning and in comparison (or in combination) with other mitigation approaches such as handwashing. The wearing of face coverings will also reduce droplet deposition on the workspace. 

 Disinfectants based on alcohols (ethanol, 2-propanol, 2-propanol with 1-propanol) as well as other common disinfectants (glutardialdehyde, formaldehyde, and povidone iodine (0.23\%–7.5\%)) are very effective against SARS-CoV-2 and come highly recommended \citep{Kampf2020}. UK Government advice to use soap and water to clean surfaces may not be the most effective since soap and water alone was not shown to deactivate the virus after 5 minutes, however if accompanied by scrubbing this may be more effective at its physical removal \citep{Kampf2020}. Sodium hypochlorite requires a concentration of at least 0.2\%, whilst hydrogen peroxide requires a concentration of at least 0.5\% and  must be left incubating for at least 1 minute. 


\subsection{Social distancing indoors} \label{sec:soc_dist}

Social distancing describes the effort to ensure that individuals remain separated by a particular distance and is often recommended primarily to reduce the transmission of disease via the droplet route. The quantification of an appropriate social separation/distance to avoid droplet transmission is often based on research by \citet{wells_air-borne_1934}. His model for disease transmission considered droplets produced by sneezing or coughing to behave ballistically, with no interaction between them; i.e. droplets would fall from the height they were produced to the floor ($\sim 2$\,m vertically), while simultaneously evaporating. In spite of the evaporation process, so called `large' droplets would reach the floor; meanwhile so called `small' droplets would evaporate quickly leaving relatively `dry' aerosol particles known as droplet nuclei. Wells proposed two mechanisms for infection: droplet transmission due to large droplets and airborne transmission due to small droplets that evaporate sufficiently to become suspended in the air for long times. Wells calculated a cut-off between small and large droplets as 100 $\mu$m (not 5$\mu$m, as is often cited).

Wells' falling-evaporation curve has been used to propose a social distancing rule by considering how far large droplets travel horizontally as they fall. The total distance travelled by a droplet is determined by its initial horizontal velocity, and also whether it is contained in a coherent flow structure caused, for example, by coughing or sneezing  \citep{bourouiba2020turbulent}. Small droplets fall more slowly than large droplets, so travel further, but they take less time to evaporate. The size of the largest droplet that totally evaporates before falling 2\,m is identified, then the horizontal distance this droplet travels is calculated and used to define a social distancing rule. For example, for droplets expelled at 10 m/s (typical for coughing) this distance is 2 m, while droplets expelled at 1 m/s (breathing) this distance is less than 1 m \citep{Xie2007}.

In addition to the work of Wells, the experiments by \citet{jennison_m_w_atomizing_1941} have also been used as evidence for social distancing of 1-2 m. Jennison used high-speed photography to examine the fall of droplets produced by talking, coughing and sneezing, concluding that the majority of the droplets fell to the ground within 1 m (the field of view for the experiments). However, no details were provided about how this conclusion was reached and it was acknowledged that the experimental method was not sensitive enough to capture all the droplets, tending to select for larger droplets \citep{bahl_airborne_2020}.

The Wells model and many of its extensions assume that droplets behave independently of each other, travelling ballistically. However, more recent research has shown that this is often not a good assumption. Experimental studies of coughing and sneezing show that exhalation results in a puff of warm, humid air that influences the distance travelled by groups of droplets \citep{bahl_airborne_2020}. Experimental images show that the turbulent gas cloud emitted from a human sneeze can travel 8\,m, carrying particles along with it \citep{bourouiba2020turbulent}. These results suggest that for coughing and sneezing, the exhalation or puff needs to be taken into account in calculating the maximum distance travelled by droplets. Moreover, consideration of these images highlights that violent respiratory events have significant directionality and in contexts where the direction of these events can be inferred then this should influence the layout of desks, etc., within classrooms and offices.

Current social distancing advice aims to reduce droplet transmission, however social distancing can also reduce transmission by small droplets, as aerosols are diluted with distance from the source \citep{chen_short-range_2020}. The suggested distance of 2 m is based on a model that assumes ballistic droplets, rather than particles that travel within an exhaled puff that dilutes with distance. Further research is needed to identify an appropriate distance at which sufficient dilution has occurred. This distance will be influenced many factors, including the nature of the exhalation (breathing, talking, coughing, etc.), the properties of the air (humidity, temperature), the droplet size distribution, the virus concentration with droplet size, and the size of the infectious dose. 

There is increasing evidence that airborne transmission of SARS-CoV-2 is significant \citep[e.g.][]{Morawska20,fennelly_particle_2020}. Outdoors, in well-ventilated indoor spaces, and for short interaction times, social distancing will reduce airborne transmission. However, for longer exposure times and/or in poorly ventilated spaces, social distancing is unlikely to be sufficient as the dilution at room scale will not reduce the aerosol concentration enough to avoid an infectious dose. This emphasises the importance of additional measures, such as good ventilation and face coverings.

\subsection{Guidance on face masks and coverings} \label{sec:masks}

There is substantial evidence that face masks and coverings can lessen the spread of COVID-19 by reducing the emissions of virus-laden particles, both droplets and aerosols \citep{leungrespiratory} from the wearer. Moreover, it has also been shown that face masks and coverings may also be protective by reducing the dose of SARS-CoV-2 received by the wearer. Face shields might prevent the escape of droplets from the user, as well as similarly protecting the user and providing some limited protection to the eyes, but they are unlikely to stop the transmission of aerosols (the airborne route) unless used in conjunction with a face mask or covering. The collated evidence presented within this section has been gathered from studies across the world and represents the most recent understanding of the subject, as presented by a  cross-section of peer-reviewed papers and scholarly works. 

In short, the use of face masks and coverings can reduce the spread of COVID-19 indoors; this is especially pertinent in settings where there is decreased physical distancing: such as shops, public transport or in work environments. However, the impact of wearing face masks or coverings on the intended interactions and activities should be carefully considered with members of disadvantaged or vulnerable groups given due attention.

\subsubsection{A summary of the evidence base for the use of face masks and coverings}

A systematic review and meta-analysis of distancing, face masks and eye wear in Lancet from June 2020 concluded that ``face mask use could result in a large reduction in risk of infection, with stronger associations with N95 or similar respirators compared with disposable surgical masks or similar \citep{chu2020physical}''. They also noted that ``transmission of viruses was lower with physical distancing of 1\,m or more'' and ``eye protection also was associated with less infection''. These reviews were in both health-care and non health-care settings. A more recent meta-analysis from August presented significant results, ``that face masks protect populations from infections and do not pose a significant risk to users'' \citep{ollila2020face}.

Masks often refer to surgical or respiratory masks (respirators) that medical staff use, whereas face coverings encompass broader types and materials such as homemade cloth masks, but may include just a simple scarf \citep{TheRoyalSociety2020new}.

 \cite{leungrespiratory} ``Demonstrated the efficacy of surgical masks to reduce coronavirus detection and viral copies in large respiratory droplets and in aerosols". Their results suggest that these masks could prevent transmission of viruses from symptomatic (and for COVID-19 pre-symptomatic/asymptomatic) individuals.


There is evidence that face masks and coverings may be effective at reducing COVID-19 cases across the world. \citet{mitze2020face} stated ``after face masks were introduced on 6 April 2020, the number of new infections fell almost to zero'' in the city of Jena, Germany; concluding ``that the daily growth rate of COVID-19 cases in the synthetic control group falls by around 40 percent due to mandatory mask-wearing relative to the control group''. Similarly, ``from epidemiological data, places that have been most effective in reducing the spread of COVID-19 have implemented universal masking, including Taiwan, Japan, Hong Kong, Singapore, and South Korea'' \citep{prather2020reducing}. 

``Our analysis reveals that the difference with and without mandated face covering represents the determinant in shaping the trends of the pandemic worldwide. \citet{zhang2020identifying} conclude that wearing of face masks (or coverings) in public corresponds to the most effective means to prevent inter-human transmission, and this inexpensive practice, in conjunction with extensive testing, quarantine, and contact tracking, poses the most probable fighting opportunity to stop the COVID-19 pandemic, prior to the development of a vaccine.'' 

“The benefits face masks could offer as a non pharmaceutical intervention were investigated using mathematical models and show that face mask use by the public could make a major contribution to reducing the impact of the COVID-19 pandemic'' \citet{stutt2020modelling}. They demonstrated that mask (and face covering) wearing can reduce transmission, with high rates of adoption likely to reduce the $R_0$ value to below one.

As of the 5$^{\textrm{th}}$ June 2020 the \citet{world2020advice} has recommended the wearing of face masks and coverings for communities and circumstances, where there is risk of transmission and in areas where physical distancing cannot be achieved, for the ``potential benefit for source control'' of COVID-19 virus. More recently (in August 2020) the WHO have released a video to advise the wearing of face masks or coverings.

\citet{TheRoyalSociety2020} DELVE Initiative reported that ``asymptomatic (including pre-symptomatic) infected individuals are infectious'' and ``respiratory droplets from infected individuals are a major mode of transmission''. Reporting that masks reduced droplet dispersal and ``cloth-based face masks reduce emission of particles by variable amounts'', similar to the percentage reductions reported (of viral, bacterial and dust particles) by surgical masks. In a more recent study, key findings were that cloth face coverings are effective in protecting the wearer and those around them and that face masks and coverings are part of ‘policy packages’ that need to be seen together with other measures such as social distancing and hand hygiene  \citep{TheRoyalSociety2020new}. 

The physics of particle capture by the materials of face masks is complex, it also recognised these mechanisms will be at play in face coverings. However, coverings (and face masks), as well as reducing the airflow which can distribute the virus, may simply prevent transmission by stopping the evaporation of droplets their escape to become droplet nuclei \citep{Leung2020}.

Face masks and coverings may also protect the wearer, to different degrees of effectiveness. The material and the make-up of the face mask or covering is important for the filtration efficacy. \citet{wilson2020covid} modelled the risk of transmission based on published data on the effectiveness of various material against a viral challenge; these included FFP2 (N95) respirator material (at 95\% efficiency), with surgical mask material slightly less efficient. A vacuum cleaner bag (83\%) was found to be the most efficient household material and a scarf (44\%) the least efficient. In between these two materials were a tea towel, cotton mix, linen, a pillowcase, silk and 100\% cotton T-shirt.

User protection is discussed by \citet{gandhi2020masks} who hypothesise that ``universal masking reduces the `inoculum' or dose of the virus for the mask-wearer, leading to more mild and asymptomatic infection manifestations.'' They state numerous cases where universal masking led to fewer cases, or more asymptomatic cases as opposed to comparative examples, whether this was in animals, on cruise ships, meat factories, or regions of universal mask wearing. ``Countries accustomed to masking since the 2003 SARS-CoV pandemic, including Japan, Hong Kong, Taiwan, Thailand, South Korea, and Singapore, and those who newly embraced masking early on in the COVID-19 pandemic, such as the Czech Republic, have fared well in terms of rates of severe illness and death."

The use of face shields as an alternative to face masks and coverings within the service industry in the UK is popular, it is also of great benefit to those of hard of hearing (HoH). ``Face shields can substantially reduce the short-term exposure of health care workers to large infectious aerosol particles, but smaller particles can remain airborne longer and flow around the face shield more easily to be inhaled." \cite{Lindsley2014}. Thus their use might prevent the escape of droplets from the user, as well as similarly protecting the user and providing some limited protection to the eyes. They are however, unlikely to stop the transmission of aerosols (the airborne route), without being used in conjunction with a face mask or covering. \citet{verma2020visualizing} visualised the flow around a face shield (and respirators with valves) and noted ``that although face shields block the initial forward motion of the jet, the expelled droplets can move around the visor with relative ease and spread out over a large area.'' Similar results were found for respirators with exhale valves, with aerosols escaping through the valve.

Any covering of the face negatively impacts the hard of hearing (HoH), with close to 11 million people in the UK who are HoH (around 1 in 6 people). Where possible, it would be prudent to adapt the guidance on face masks to accommodate them, as the usage of face masks may hinder the ability to listen and lip-read. An effective strategy to accommodate those HoH is to use clear face masks~\citep{grote2020covid}, or the use of novel technologies such as captioning apps.

Finally, the application of the precautionary principle suggests that people should be encouraged to wear face masks and coverings on the grounds that, in many circumstances, we have little to lose and potentially something to gain from this measure. \citet{greenhalgh2020face} said that ``masks are simple, cheap, and potentially effective \ldots and outside the home in situations where meeting others is likely (for example, shopping, public transport), they could have a substantial impact on transmission with a relatively small impact on social and economic life.''



\subsection{Source reduction through timetabling and purging between events} \label{sec:purge}

Airborne infection risk is reduced when the ventilation provision of outdoor air is maximised. Operating the existing indoor environment conditioning and controlling equipment in a manner that fixes the outdoor air supply rate to be maximal (with due consideration to the practical limits for a comfortable indoor environment), the airborne risk can be greatly reduced by lowering occupancy in a given indoor space. For example, should the ventilation plant be kept running at the same level (i.e. unchanged absolute outdoor air supply rate) and the occupancy halved (e.g. through week in -- week out working) in an indoor space then the chances that infection occurs within is approximately halved. 

Where reductions to the absolute levels of occupancy are impractical then some change of timetabling should be considered. Where possible, attendance should be extended by some occupants arriving and leaving earlier than usual with other occupants arriving and leaving later than usual. Moreover, consideration should be given to purging rooms between meetings, classes and events. This would require the room to be unoccupied between consecutive events during which period all possible efforts are made to increase the outdoor air supply rate (whether by opening windows, doors and ventilation systems). At the end of the purging period it is best if the room is cleaned in the manner detailed in \S\ref{sec:contact} to minimise the chance of spread to the next occupants. It is highly likely that the greatest rates of decay in the concentration of virus-laden particles will occur at the start of the purging periods. So any purging duration is better than none as long as the increased ventilation flows are given time to establish themselves. That said, the longer the purging duration then the lower will be the resulting concentration levels. To establish how effective these purging strategies may be, a simple `room-change' time scale can be calculated by dividing the volume of the space by an estimate of the rate of outdoor air supply during purging; note that the room-change time scale, in hours, is simply the inverse of the air changes per hour (AC/hr). \citet{Melikov20} report the intake fraction (the proportion of air exhaled by the infected person that is then inhaled by another occupant) for various purge scenarios. They quantified the reduction in intake fraction for ventilated cases with purging periods (of between 15\,min and 30\,min) which were in the range 0.2--1.8 room-change time scales. They went to consider cases of increased ventilation rates which led to purging times (15\,min) being around 5.5 room-change time scales. Their conclusions included the suggestion that periods of constant occupation should be short with appropriately long breaks being recommended; break durations of 10--20\,min were recommended for occupancy durations of 30--45\,min for classrooms, meeting rooms, conference rooms, etc.... 

In summary, wherever possible occupancy should be reduced (by remote working and/or reduced occupancy), additionally unoccupied periods should be introduced at regular intervals throughout the day during which the space should be purged and after which the space should be cleaned. Risk-based calculations are needed to determine the optimal purging times. 


\subsection{The influence of occupancy behaviour on indoor air movement: implications for the spread of COVID-19} \label{sec:people}

The movement of people within enclosed spaces leads to considerable disturbance of the air and any airborne aerosols. Although aerosols greater than about 10 micron settle from the space relatively rapidly and so they may be dispersed by people movement, smaller aerosols (e.g. $<5\,\mu$m) which can remain airborne for several 10’s of minutes may be much more strongly affected by such dispersal. This dispersal may in fact dominate  other dispersal processes if there is a sufficient frequency of people passing through a space. People have widths typically in the range $0.3-0.5$\,m,  and move at speeds of $1-2$\,m/s even when walking, and this leads to a highly turbulent wake with  Reynolds number of about $100,000$. The mixing associated with this wake in corridors, supermarket aisles, meeting rooms, school classrooms or other spaces with relatively high people density and movement (even with $2$\,m spacing) may be  key for quantifying the aerosol dispersion prior to it being ventilated or settling out from the space. Indeed, with ventilation timescales of $10-20$\,minutes in typical buildings, with $5-10$ air changes per hour, and the settling time of small aerosols ($<$10\,$\mu$m) being of comparable duration, the aerosols may be mixed by the wakes of many people. For example, in a supermarket, with one person moving down an aisle every $10-30$\,s, a cloud of infected aerosol may be mixed by the wakes of between $20-120$ people. This mixing leads to a more uniform, but smaller concentration of aerosol in space, thereby increasing the risk of exposure to some aerosol for the subsequent people that pass by the aisle, although the amount of aerosol may be smaller; the associated risk of infection from any virus in these aerosols depends on dose and hence the amount of aerosol to which they are exposed \citep{Bhamidipati20}.

\begin{figure}
	\centering
	\includegraphics[width=\textwidth]{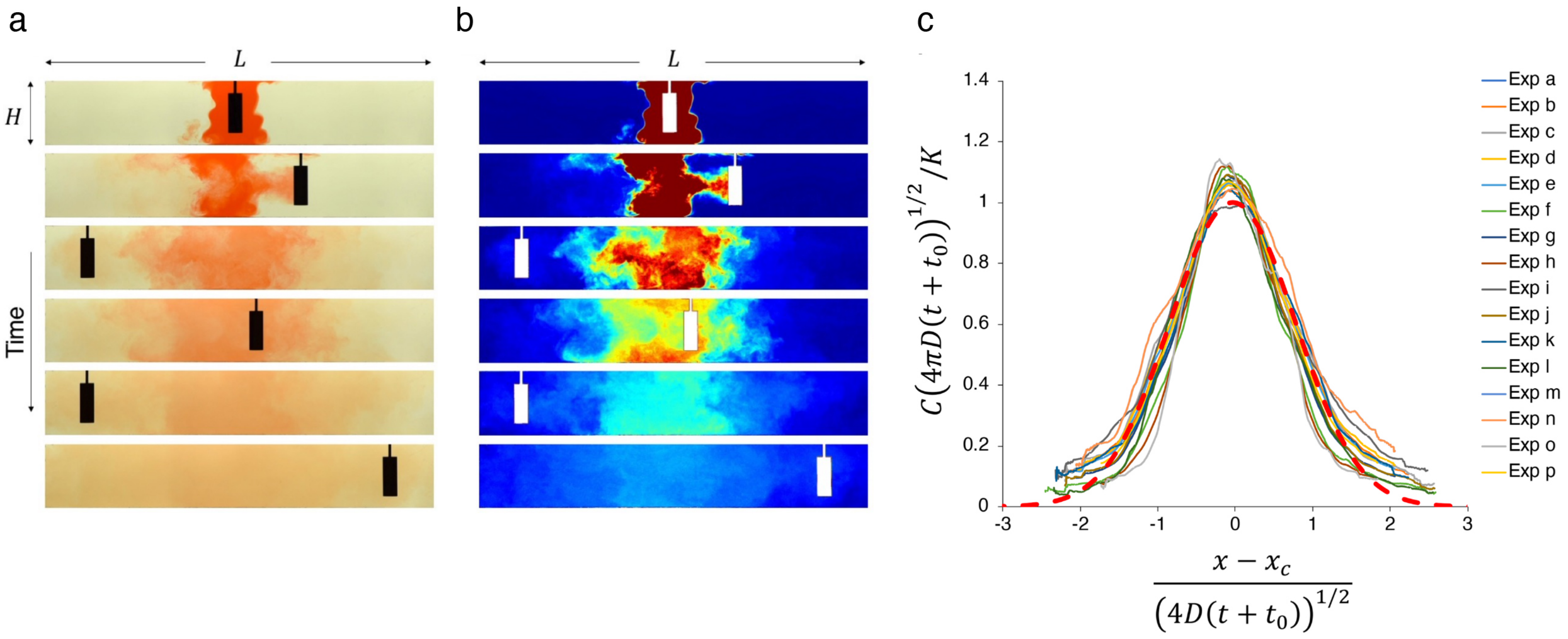}
	\caption{(a, b) Images show the depth-averaged concentration of a cloud of dye in a laboratory tank. This evolves in time owing to  the mixing produced by the repeated motion of a cylinder, representing movement of people in a corridor. In (a), pictures of the tank are shown as captured during an experiment. In (b), false colours are used to represent the dye concentration field in each of these pictures, with red being the maximum concentration and blue showing absence of dye. The white rectangle on each image in (b) illustrates the position of the cylinder at that time. (c) Collapse of the experimental data of the depth-averaged dye concentration  to a continuum model of the concentration of the dye along the channel. The $y$ axis shows the dimensionless concentration, at each time scaled relative to the theoretical  value in the centre of the channel at that time, and the $x$ axis shows the dimensionless distance along the channel, scaled with the predicted diffusive spreading along the channel at each time  \citep[after][]{Mingotti20}.}
	\label{AWfig_01}
\end{figure}

Laboratory simulations of the dispersal of both clouds of dye and suspended particles have been carried out in a fluid-filled channel of size $1.04$ m $\times$ 0.10\,m $\times$ 0.20\,m, as a model of a corridor. To model the movement of people, cylinders of radius $0.015-0.050$\,m were moved back and forth along the channel, with speeds of $0.1-0.2$\,m/s, thereby providing a dynamically similar flow regime for the full-scale flow of people's wakes. This models the mixing of the dye or cloud of particles along the channel. Data shows that the motion of the cylinder leads to an effective dispersion coefficient for the along-channel mixing, which provides the basis for a theoretical model \citep{Mingotti20}. In experiments with a background ventilation along the corridor, in addition to the people-driven mixing, a dilution wave migrates along the corridor after the release of infected aerosol along the corridor, but the dispersion associated with people movement causes the aerosol to mix back upstream into the dilution wave, delaying the effectiveness of the ventilation \citep{mw20}. In figure \ref{AWfig_01}(a, b), images at successive times from an experiment illustrate the dispersal of a cloud of dye along the channel. In figure \ref{AWfig_01}(c), the concentration data from a number of experiments collapse to a simple model for the dispersion. 

Scaling up to a building, we find that the typical mixing rates associated with people in a corridor/supermarket aisle depend on the frequency of passage of people. With a person walking along the corridor/aisle every $10-40$\,s \citep{Choudhary2010, Wijk2018}, this corresponds to a dispersion coefficient in the range $0.05-0.2$\,m$^2$/s \citep{Mingotti20}, so that over a period of $600-1200$\,s, airborne aerosol may spread about $5-15$\,m along the corridor.

\begin{figure}
\centering
  \includegraphics[width=0.5\textwidth]{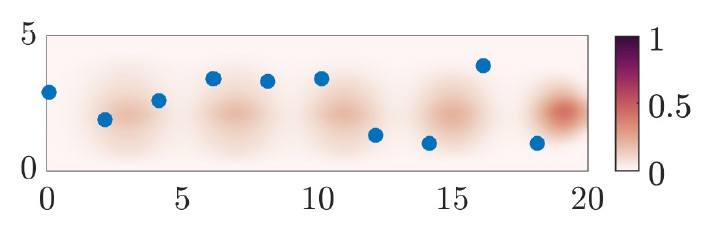}
  \caption{Image showing the mixing of individual clouds of infected aerosol, dyed different shades of red, and normalised relative to the initial concentration as seen on the legend. The corridor is 5 m wide ($y$ axis) and 20 m long ($x$ axis), with people (blue dots) moving from left to right along the corridor. In this simulation, the along-corridor people spacing is 2 m and they move with speed 1.5 m/s, while the clouds of aerosol are produced by one person moving down the corridor, so the older cloud at the left hand end of the corridor is more dispersed than that on the right (after \citet{Bhamidipati20}).}
\label{AWfig_02}
\end{figure}

\citet{Bhamidipati20} developed a simple computational model for the dispersal of aerosols in a building by a stream of individual people moving through a building, modelling the detailed mixing produced by each person. This has led to a series of simulations of the dispersal of clouds of aerosol as people pass along a corridor. Figure \ref{AWfig_02} shows the concentration of aerosol with time, over a period of 15\,s as an infected person walks down a corridor breathing out. A series of local clouds of aerosol-laden air are breathed out by the infected person, and these are then dispersed owing to the mixing produced by the continuing stream of people following in the wake of the infected person. 

Models of different building types are under continued development, but the key result is the efficacy of mixing produced by the movement of people, which can lead to widespread dispersal of small aerosol produced by an infected person. This has leading order implications for the occupancy levels in shops and in the corridors of classrooms and offices in terms of the risks of exposure to the small aerosols due to the repeated passage of people. The combination of aerosol settling and ventilation of the air from the space typically leads to a residence time of the small aerosols of several tens of minutes; if these small aerosols are present in sufficient numbers to play a role in infection transmission, which depends on the source strength (i.e. the number of infectious people present), and the residence time (i.e. as regulated by the ventilation rate), then the continued presence of the infectious and healthy people may provide a pathway for transmission. 

Further experiments have been carried out in several operational buildings in which localised dilute clouds of CO$_2$ are released at a point in a room or corridor, and the subsequent spreading of this cloud is then measured over time; the CO$_2$ acts as an analogue for the small aerosols in that it moves with the air flow in the space \citep{wgm20}. Comparisons have been made between the rate of dilution and flushing of the CO$_2$ from the space in the case with people moving and with no people moving in the space. In the example of a ventilated corridor, for example, the impact of the people moving along the corridor is to drive additional mixing of the cloud of CO$_2$ into the new ventilation air, thus delaying the flushing  of the CO$_2$ from the space. This illustrates how the internal mixing processes of the air in buildings, and especially people movement, can lead to greater residence times for small aerosols \citep{wgm20}. To mitigate the risks of these small aerosols, potential solutions include increasing ventilation rates, reducing the duration of exposure, and reducing the source of the aerosols by reducing occupancy levels and through the use of face masks. 


\section{Ventilation and the airborne transmission route} \label{sec:airb}

The adequate ventilation of a building space should be regarded as the primary mitigating measure against the spread of airborne diseases. In temperate climates this leads to the simple advice that all ventilation (by which we refer exclusively to the supply of outdoor, or suitably sterilised or filtered, air) systems be operated to maximise supply and ventilation openings (e.g.\ windows, vents, louvres, doors, etc.) be opened to the extent permitted by design. 
However, in colder periods (like the British heating season) there is a conflict between reducing airborne infection risk, by increasing the outdoor air supply, and maintaining occupants thermal comfort and reducing energy consumption and the associated costs. In this section we seek to provide guidance to resolve this conflict.

In order to precisely quantify the risk of airborne infection occurring it is necessary to determine occupants' exposure to airborne virus particles. To do so rigorously is challenging (often to the point of impracticality see \S\ref{sec:aibUnder}) and thus it is wise to focus efforts on estimating the likelihood, perhaps better the change in likelihood, that airborne infection may occur within a space under a number of scenarios under the assumptions made.

\subsection{Towards an understanding of ventilation and COVID-19} \label{sec:aibUnder}

Ventilation, i.e.\ the supply of outdoor air, dilutes any pollutants produced indoors --- including viable airborne virus particles. This dilution is dominated either: by the incoming outdoor air mixing with the existing indoor air, with then further mixing occurring as the air is transported through the building space before the ultimate evacuation outdoors; or by seeking to introduce the outdoor air into the occupied zone of the building space in a relatively unmixed state and then `displace' any air polluted by airborne virus particles back outdoors \citep[e.g.][]{bl08, bl08a, Mingotti15b, Mingotti15a}. Irrespective of the intended strategy, mixing will occur inhomogeneously within the building space which results in unpredictable distributions of virus particles. Moreover, differences in temperature between indoor and outdoor air and the production of heat within the building space (every occupant outputting $\sim$100\,W of heat) exacerbate the complexity of indoor air flows and increase the unpredictability of the distribution of airborne virus particles. Finally, in all indoor spaces (perhaps excluding COVID-19 hospital wards) it is unknown if there are any infectious occupants (the sources) and where they might be located. As such, trying to predict the distribution of viral contamination, for example using normal exhaust ventilation design techniques (which require knowledge of the source location and rate of contamination) is likely to be futile.
Therefore estimating the risk of infection via modelled data typically requires the assumption that the distribution of airborne virus laden particles is relatively uniform irrespective of reality. We note that in the case that `dead' or `stagnant' zones can be expected or evidenced within an indoor space then further considerations are required see, for example \S\ref{sec:other} for further discussion.

As such, the critical input pertaining to the control of the environmental quality for any model estimating airborne infection risk within an indoor space is the bulk supply of outdoor air, or ventilation rate per person which we denote $Q$. However, outdoor air may enter an indoor space via a ventilation system, windows, doors, vents, cracks in the building fabric or, indeed, though the very fabric itself (i.e.\ many building materials, e.g.\ bricks, are porous). As such, there is significant scope for both intentional and unintentional supply of outdoor air. Directly measuring the air flow through all of the potential pathways for any given indoor space is challenging. Pressure testing can be used to measure infiltration rates but cannot assess the ventilation rates in operational settings and does not help with individual spaces or zones. For purely mechanically ventilated, well-sealed, buildings it may be tempting to assume the intended outdoor air supply is actually achieved; however, such an assumption is not without risk. Measurements of the actual air flows could be made in the duct work but these are not without their own challenges (the velocity profiles of air flows in the these typically tortuous ducts is a challenging area of fluid mechanics in its own right) .

In most indoor environments the dominant source of \co is human activity and the level of \co in outdoor air remains broadly constant (at about 400\,ppm). As such, it may be tempting to try to infer outdoor air supply (ventilation) rates by monitoring \co\!. However, whilst \co is an excellent proxy by which to determine indicative levels of ventilation, rigorously determining the precise (typically transient) outdoor air supply rate is non-trivial \citep[e.g. see the appendices to][]{Burridge20}. 

Crucially, \citet{rm03} established a methodology, based on the Wells-Riley model \citep{Wells55,Riley78}, which takes monitored \co data (the equipment to do so costing less than a couple of hundred pounds per sensor) and directly infers the risk of airborne infection without the need to assess/assume the ventilation rate (nor does it require the space to be in steady-state). Their model, assuming the presence of an infector, can be parameterised for any airborne disease to provide both the likelihood of infection and the reproductive number (or R-number, i.e. the average number of infections arising from a single infectious case) for indoor spaces over time periods selected such that the space remains constantly occupied. 

In addition to the monitored \co, the risk reported by \citet{rm03} depends strongly on the time period of assessment, the occupancy level, the nature of the virus, and occupancy activity --- the latter two aspects being parameterised via the rate of generation of infectious quanta $q$ per person (usually expressed in quanta per hour). \citet{Wells55} conceived the idea of a quantum (or infectious dose) in an effort to describe the stochastic behavior of airborne infection, and values for the quanta generation rate have been derived for SARS-CoV-2 \citep[e.g.][]{Buonanno20}. However, just as for most other airborne diseases, there is wide variation in the values relevant for an infectious individual depending on a) their activity level, b) the viral load in their sputum and c) the ratio between one infectious quantum and the infectious dose. 

\citet{Buonanno20} report that in typical scenarios with low activity levels, $q \approx 1$\,/hr may be appropriate for COVID-19. \citet{Buonanno20} further suggest that should an individual be vocalising (in a manner not dissimilar to talking) whilst carrying out light exercise (e.g. walking) then values as high as $q \approx 100$\,hr can be inferred --- for offices occupied and ventilated at appropriate levels \citep[e.g. $\sim 10$\,m$^2$ per occupant and the outdoor air supply per person is $Q \approx 10$\,l/s/p][]{CIBSEA} then for occupancy periods of constant occupancy of, say, four hours the R-number for the office (assuming a floor-to-ceiling height of between 3--4\,m) would be around 4, i.e. from a single infector then four new COVID-19 infections would result. For more typical behaviour in an office, i.e. most occupants carrying out desk work with a few talking relatively quietly for which $q \approx 1$\,/hr, then even assuming the office remains constantly occupied for a full nine hour day, R-numbers of around 0.1 are obtained (these rise to around 0.2 if the office is poorly ventilated, i.e.\ $Q=4$\,l/s/p, and drop to around 0.06 if the ventilation is doubled). 

The results of \citet{Buonanno20} take a value of $q=142$\,/hr and show the airborne infection risk for various public indoor spaces (namely, shops and restaurants), reporting R-numbers for differing exposure scenarios (changing outdoor air supply and occupancy scenarios) which they describe as before and after `lockdown' (N.B. they do not report values for restaurants after lockdown). They select modelled durations of $\sim$ 3\,hrs. For poorly ventilated spaces \citet{Buonanno20} report R-numbers ranging from $2 \leq R \leq 50$ before lockdown ($1.7\,\textrm{l/s/p} \geq Q \geq $ 0.2\,\textrm{l/s/p}) and after lockdown $0.1 \leq R \leq 0.8$ after ($10.5\,\textrm{l/s/p} \geq Q \geq $ 5.2\,\textrm{l/s/p}). For better ventilated spaces then $1 \leq R \leq 6$ ($10\,\textrm{l/s/p} \geq Q \geq $ 4.5\,\textrm{l/s/p}) before lockdown and afterwards $0.1 \leq R \leq 0.4$ ($55\,\textrm{l/s/p} \geq Q \geq $ 22\,\textrm{l/s/p}). 

Since an occupant can become infected at any point on any given day then taking any particular duration seems and arbitrary choice which, for the most part, is hard to justify. \citet{Burridge20} point out that for indoor spaces which are regularly/consistently attended by the same/similar group of people (e.g. open-plan offices or some school classrooms, and herein a `regularly attended space') one should consider the likelihood of infection over a period during which an infectious person may remain pre/asymptomatic. For COVID-19 this period is currently estimated to be between five and seven days. They developed simple extensions to the work of \citet{rm03} which enabled variations in occupancy behaviour and activity to be accounted for and the likelihood of infection to be assessed from monitored \co\!. In doing so, \citet{Burridge20} calculate the likelihood that an indoor space contributes to the spread of COVID-19 by assuming that a single pre/asymptomatic infector regularly attends the space and ceases to do so once they show symptoms reporting the absolute R-number, $R_A$, for modelled and monitored spaces. 

\begin{table}[]
    \centering
    \begin{tabular}{l|c |c |c}
         R-numbers, $R_A$ & \; $Q=4$\,l/s/p \; & \; $Q=10$\,l/s/p \; & \; $Q=20$\,l/s/p \;   \\
         \hline
         $q=0.3$\,/hr & 0.25 & 0.13 & 0.07 \\
         $q=1$\,/hr & 0.84 & 0.42 & 0.24  \\
         $q=5$\,/hr & 4.0 & 2.1 & 1.2  \\
         \\
         $q=20$\,/hr & 14 & 7.6 & 4.4  \\
         $q=100$\,/hr & 35 & 26 & 18  
    \end{tabular}
    \caption{COVID-19 R-numbers, $R_A$, calculated over the period that a pre/asymptomatic person remains attending work in an open-place office (floor plan of 400\,m$^2$ and floor-to-ceiling height of 3.5\,m) occupied by 40 people for 8\,hrs each day \citep[see][]{Burridge20}.}
    \label{tab:R}
\end{table}

As shown in table \ref{tab:R} for typical behaviour in a typical office \citet{Burridge20} report $R_A \approx 0.5$ which rises to $R_A \approx 0.8$ for a poorly ventilated space ($Q=0.4\,$l/s/p) and falls to $R_A \approx 0.2$ if the ventilation is doubled --- which are reassuringly below one, thereby suggesting that the return to desk-based office work is unlikely to contribute significantly to the COVID-19 spread. However, should most of the office remain sedentary but start vocalising (akin to a call-center, or similar) then $R_A \approx 2$ --- highlighting the importance of occupancy behaviour in determining whether or not a particular indoor space contributes to the spread of COVID-19 or not. \citet{Burridge20} also present results for the airborne infection risk from \co data monitored within an office with uncontrolled natural ventilation. The particular office is not of a modern well-sealed design. The risk levels within that monitored naturally ventilated office remain comparable with the modelled office, intended as being `typical', therein. However, the airborne infection risk was calculated for periods when the windows were opened and for those when the windows remained shut --- the risk of infection was approximately doubled when the windows remained shut. This provides quantitative support for the guidance that where practical ventilation openings (like windows, doors, etc.) should be opened, but doing so is not without conflict, see appendix \ref{app:NV}.

\begin{figure}
             \begin{center}
             \includegraphics[width=0.7\textwidth]{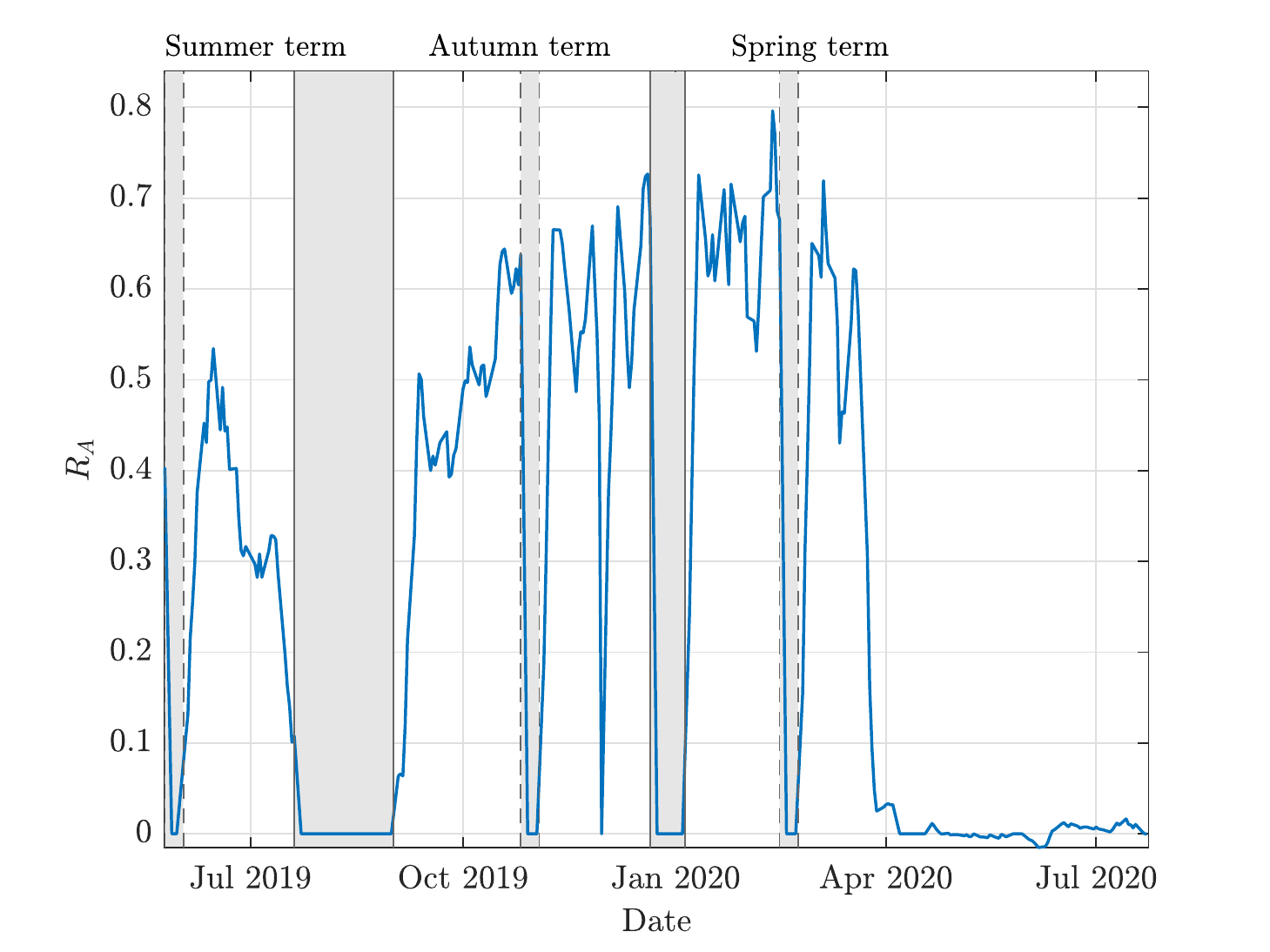} 
             \end{center}
             \caption{The variation in the absolute R-number, $R_A$, during the period 25$^{\textrm{th}}$ May 2019 to 20$^{\textrm{th}}$ July 2020 determined from monitored \co and the school's occupancy timetable for a typical classroom within a relatively modern build school (rebuilt in September 2016). The shaded regions correspond to holiday periods. For data recorded during the summer term of 2019 the average was $R_A = 0.35$, for the autumn term of 2020 the average was $R_A = 0.53$, and for data recorded during the spring term of 2020 the average was $R_A = 0.63$ until the lockdown of 20$^{\textrm{th}}$ March.}\label{fig:Vour}
\end{figure}

Such findings beg the question: how might the infection risk vary between seasons as weather changes from being typically temperate and what might this mean for the spread of COVID-19 in winter 2020? \citet{Vouriot20} examined the R-number within various spaces inside schools. Their most meaningful results are for classrooms, which can be regarded as regularly attended spaces, where they reassuringly find R-numbers below one if pupils are assumed to be carrying out desk based learning in a relatively calm/quiet environment and quanta generation of $q \approx 1$ are appropriate. The R-numbers rise to $1.5 \leq R_A \leq 3.8$ if one assumes values of $q \approx 5$ which are more appropriate if the class is carrying out more vocal activities and they rise further still if one assume pupils are actively moving around the classroom. Crucially, in all cases, the greatest R-numbers are obtained during colder winter months (for example, November to February) --- being typically around 80\% greater than those estimated for more temperate months (i.e. May to September). We see no reason not to expect similar trends in the reproductive number of indoor spaces beyond schools and such convictions underpin our guidance to assess and ideally monitor ventilation provision in order to understand the needs for modification and the implementation of other measures. 
 
\subsection{Ventilation guidance for winter 2020} \label{sec:VentGuide}

A concern for the northern hemisphere winter is that buildings will become less well ventilated, with a lower supply of outside air in order to maintain warm conditions indoors. Ventilation systems that recirculate a proportion of the indoor air, primarily to temper the temperature of the outdoor air without increasing energy consumption, are common but it is crucial to regard only the flow of outdoor air as contributing to the ventilation rate (see appendix \ref{sec:InDuct} for more detailed discussion). We recommend evaluating the benefits of increased monitoring of the indoor environment, especially indicative outdoor air supply rates via \co levels. \co levels not exceeding 1000\,ppm within an indoor space broadly indicating that outdoor air supply is likely to be adequate for offices and mechanically ventilated classrooms (i.e. $Q \approx 10$\,l/s/p), with the equivalent level being 1\,500\,ppm in naturally ventilated school classrooms \citep[i.e. $Q \approx 5$\,l/s/p, see Annex A of][for a full discussion]{BB101}. Doing so will assist in: a) highlighting high-risk spaces for which mitigation measures need to be considered (e.g. enhanced ventilation provision to manage certain activities or some of those measures detailed in \S\ref{sec:other}), and b) helping obtain the quantitative evidence to reassure occupants as to their relative safety. Where monitoring of \co is undertaken, consideration of the build-up and decay of \co, and the variation in \co levels within the space can provide additional insights; where experience or skills are lacking, engagement with a professional building services engineer would be beneficial. We note that the design of school classrooms will, in some cases, have been carried out allowing \co levels to reach up to 2\,000\,ppm for brief periods; as \citet{Vouriot20} show, this does not necessarily indicate unacceptable risk levels for COVID-19. Moreover, active management of a classroom's windows and other ventilation openings may enable these high peaks to be avoided or minimised, appendix \ref{app:NV} provides discussion of strategies to do so in winter without excessively impacting occupant thermal comfort.

Where people are brought together in moderate to high densities for significant portions of the day and no monitoring is intended, we recommend that the design provision for ventilation of the indoor space be investigated to determine whether adequate ventilation is likely being achieved. Consideration should be given to installation and the subsequent maintenance to estimate whether or not the design provision might realistically be attained. Where the design provision is unavailable, efforts should be made to `reverse engineer' an understanding of the ventilation. In the absence of knowledge as to the number of infected occupants, ventilation provision should always be considered per capita based on the design occupancy --- we ventilate buildings for the sake of the occupants. Decreasing occupancy density should be considered both to minimise transmission via the droplet route, the fomite route (see \ref{sec:soc_dist}) and, crucially, the airborne route as the per capita ventilation provision from fixed systems/plant can be drastically increased without new investment (see \ref{sec:purge}). Where ventilation rates meet existing design guidance the expected risk of airborne infection might be regarded as low. In the absence of monitoring, consideration should be given to potential stagnant zones within indoor space (e.g. a sheltered reading corners or break-out spaces); where suspected these should be addressed (see \ref{sec:other}).

Where \co monitoring is carried out, sensors should be placed at heights and locations within the space representative of the breathing zone. 
Where practical, occupancy and excess \co (i.e. \co above outdoor levels) should be recorded and simple calculations undertaken to estimate the level of risk to occupants \citep[see, e.g.][]{Burridge20} --- we expect that, in the most part, these calculations will prove simply reassuring. Where no calculations are desired, then for given activity within a space, it is worth noting that the airborne infection risk depends directly on the excess \co relative to the number of persons responsible for producing it --- this latter quantity being impossible to evaluate without monitoring (in which case the simple calculations of \citet{Burridge20} may prove more useful). In all cases, we suggest that consideration not only of risk but also the rate at which the risk is increasing with time is prudent. The relative rate of increase in airborne infection risk is directly proportional to the ratio of the instantaneous values excess \co and current occupancy \citep[see][]{Burridge20}. These values are easily obtained and indicate the level of risk that will be realised should all else remain equal. 

We note, however, that ventilation flows are complex and may involve air flow from one space in a building to another. They are also strongly influenced by the locations of openings and the strength of internal heat gains. Further work on the implications of these effects is needed.

Finally, we conclude that the risk of COVID-19 being spread by the airborne route is not insignificant, varies widely with activity level and environmental conditions (which are predominantly determined by the bulk supply of outdoor air), and are expected to increase in winter 2020 relative to summer. 


\section{Other mitigating measures} \label{sec:other}


This section explores some of the additional measures available to mitigate risk of airborne transmission of COVID-19 in indoor environments. Here by `airborne' transmission we mean transmission via smaller particles which can be suspended in the air for a considerable time. By `droplet' transmission we mean short-range transmission via larger droplets which fall to a surface within seconds and within a few meters of the infected person.

The risk of airborne transmission indoors can be mitigated through dilution of the indoor air by clean outdoor air, as discussed in \S\ref{sec:airb}. This requires a ventilation system for which the air intake can be increased, or installing secondary ventilation systems. Substantially increasing the ventilation in a space is often impractical. 
Further, in order to reduce the risk of infection by a factor, the ventilation rate must be increased by the equivalent factor. Increasing the volume of outdoor air becomes particularly challenging in winter without compromising the thermal comfort of occupants or energy use (due to an increased heating load). These factors limit the viability of reducing risk by increasing ventilation.

The mechanisms for the transmission of COVID-19 are not yet well understood. Reducing the risk of infection to zero is not possible (short of global eradication), therefore, where practical to do so, all available measures should be taken in order to gain the biggest reduction in risk. In the absence of or to complement increased ventilation, alternative engineering control measures can be used. Some of the measures available are discussed here.

%


\subsection{Air filters or cleaners}

Filters and air cleaners, such as UV cleaners or photocatalytic oxidation, can either be installed directly within the existing ventilation system, removing virus-laden particles from recirculated air, or as independent units within a room to supplement the existing ventilation. 

In their latest guidance document in response to the pandemic, The Federation of European Heating, Ventilation and Air Conditioning associations (REHVA) recommend avoiding the use of centralised recirculation as typical local air filters within these systems are not effective at filtering out viral material which tends to be too small for the filter \citep{rehva202}. Installing high-efficiency particulate air (HEPA) filters would allow virus-laden particles to be removed, however these are not easily installed in existing systems and further system modifications are required in order to provide a higher pressure drop to maintain the same airflow rate. 

Standalone air filters can be effective at removing virus-laden particles provided they target the appropriate range of particle size. However, regardless of the efficacy of the filter itself, the supply of clean air is limited by the flow rate of air passing through the filter. Despite this, the supply of clean air in a space can be up to 1000 $m^3/hr$ \citep{ZURAIMI20112512}. However, filters require regular servicing to maintain their effectiveness - clogged filters lead to a build up of viral particles and can act as a source of viral matter rather than a sink \citep{Eames2009}. 

Air cleaners such as UV air cleaners are also limited by the volume of air that can be passed through the device. UV light has been shown to be effective in deactivating various viruses under laboratory conditions, including coronaviruses \citep{Morawska20,Beggs2020}. While the evidence that UV is effective against SARS-CoV-2 specifically is currently limited, it seems highly likely to be the case \citep{Beggs2020}. 

The efficacy of air filters and cleaners is likely to be highly sensitive to their location within a room. Depending on the size, shape and airflow patterns within a room, the air within some areas of the room may never reach the device. In the worst case scenario, the device simply recirculates the same small volume of air within a much larger room. Therefore while these devices tend to promise certain air changes per hour (AC/hr) equivalent of clean air, in reality this is only true if the air pulled in by the device has not already been cleaned. In smaller, poorly ventilated rooms such devices are likely to provide significant benefit, however for larger spaces an understanding of the airflow patterns within the room is required to ensure that the device is effective.

Care should be taken when considering which air cleaner/filter to use as many devices have been found to have a limited effect \citep{siegel2016,ZURAIMI20112512}. HEPA filters are often recommended as the most effective technology currently available  \citep[e.g.][]{ZURAIMI20112512}.

\subsection{Personalised ventilation}

Personalised ventilation (PV) supplies clean air directly to the breathing zone of an occupant of a room via a device installed at their workstation. A certain minimum velocity is required for the supplied air in order to penetrate the convective flow driven by body heat \citep{melikov2004}. Further, a large target area is desired in order to account for occupants' movement. The required rate of clean air supply can therefore be high. While studies have shown that PV can be effective at reducing risk for occupants while at their workstations \citep{MELIKOV2002829,melikov2004}, protection is not provided to occupants when away from their workstations. PV may in fact facilitate the transport of exhaled pathogens to other occupants \citep{Bolashikov2009}. Installing a personalised device for all occupants in an office is also likely to be expensive and impractical. 

PV may be most viable in scenarios where the occupant is required to remain at their workstation for the duration of their shift.

\subsection{Desk and ceiling fans}

While desk or ceiling fans do not enhance the bulk supply of outdoor air, they can be used to increase air mixing within a room, which may lead to a more homogeneous distribution of virus-laden particles. Where areas of stagnant air are identified (or are suspected), a fan can be used to increase mixing with the wider space, therefore potentially reducing the risk of the accumulation of virus-laden particles within the stagnant zone. 

The localised (relatively) high velocity air flows produced by desk fans may result in increased re-suspension of virus-laden particles and this risk should be considered.
However, studies have found that using either desk or ceiling fans to increase air mixing within a room can lead to increased rates of deposition \citep{Mosley2001,THATCHER20021811,Xu1994b}. These experiments use oil droplets \citep{THATCHER20021811}, cigarette smoke particles \citep{Xu1994b} and a combination of oil droplets, salt and incense \citep{Mosley2001} under controlled laboratory conditions. However, deposition rates can vary by orders of magnitude depending on the particle size, room surface-to-volume ratio (which was varied predominantly via the inclusion of furniture) and airflow speeds and turbulence within the space. Generally the removal of airborne particles through deposition is likely to be much lower than that through ventilation, however for poorly ventilated rooms these rates can be comparable, particularly for larger particles ($>1\mu m$). It is unclear whether the results of these studies would translate to increased deposition rates of COVID-19-laden particles under real-world conditions. The impact of changing deposition rates on the risk of surface transmission via fomites is also unclear. 

Due to this uncertainty and the potential risk of increasing the re-suspension of virus-laden particles within the space, the use of desk fans is generally not recommended. However, when a stagnant zone is evident, either via intuition (e.g. sheltered spaces like reading corners within classrooms and breakout spaces within open-plan offices) or through monitoring CO$_2$ concentrations, the benefit of using a desk fan to increase mixing with the wider space 
may outweigh the potential drawbacks. 
Where fans are used in this capacity they should be orientated such that increased mixing is achieved between the stagnant zone and the surrounding space (into which the outdoor air provision should be checked).

Ceiling fans promote vertical mixing of air within rooms and, in addition to increased bulk mixing, help reduce temperature stratifications from forming within the room. As such where the ventilation strategy relies on stratification within the room, e.g. where a displacement ventilation strategy can be successfully evidenced, the use of ceiling fans is not recommended. Otherwise, temperature stratification within a room could significantly inhibit the vertical mixing, and dilution, of virus-laden particles from within the breathing zone. Moreover, in the heating season the downwards mixing of warmer air from the ceiling can enable the increased supply of outdoor air without compromising thermal comfort (nor increasing the energy consumption associated with heating). Further, their use in conjunction with upper room UV (see \S\ref{sec:URUV}) has been found to greatly increase the exposure of virus-laden particles to the upper region of the room \citep{74McDevitt,Morawska20}.

\subsection{Air ionisation}

Air ionisation is a relatively new technology and involves the production of ions such as the hydroxyl radical ($OH^-$) from a corona discharge between two high potential electrodes. These ions have been shown to have germicidal properties as they react with the surface structure of the pathogen \citep{Bolashikov2009}. The technology was shown to be effective against certain bacterial pathogens when installed in a hospital intensive care unit, but with no effect on others \citep{Kerr2006}. Ionisers are easy to deploy and have high energy efficiency 
but have not been evidenced as effective devices against viruses including SARS-CoV-2. Some devices are known to produce ozone so their effect on air quality should also be considered \citep{siegel2016}. However, PV has mainly been researched in idealised laboratory settings and its potential for mitigating airborne infection is not yet understood.


\subsection{The use of screens}

Using screens to provide a physical barrier between the occupants of a space is a simple and easily applied measure to mitigate the risk of transmission. The use of screens is widespread in hospitals and the commercial sector; in supermarkets they are used to provide a barrier between the shop assistant and the customer.

However, this application is targeted at reducing the risk of infection via larger droplets. There are very few examples of the use of screens to mitigate airborne transmission (i.e. smaller droplets or droplet nuclei), either by providing a barrier directly between occupants, or in an attempt to favourably manipulate the airflow patterns within a space. \citet{69Noakes} use CFD simulations to investigate the effect of partitioning a hospital ward room on the risk of patient-to-patient and patient-to-visitor transmission. They conclude that, combined with carefully considered changes to the positioning and number of the ventilation inlets and outlets, the risk of infection can be reduced significantly. However, in some instances the risk of transmission from patient to visitor was increased by the presence of the partition due to reduced airflow in certain areas of the room. 

This highlights the main problem with the use of screens indoors; the impact of the screen on the airflow patterns within a space is very difficult to predict. While the exchange of air between two areas of a room may be reduced, the presence of a screen can lead to areas of stagnant or recirculating flow where the virus could accumulate. Further to this, installing screens which affect airflow would require an evaluation of the impact on any environmental monitoring undertaken, and crucially on fire safety (e.g. as they may prevent smoke from reaching the smoke detector or have other fire safety implications). 


\subsection{Upper room Ultra-Violet Germicidal Irradiation} \label{sec:URUV}

Upper room UV provides a way to utilise the deactivating properties of UV light without the limitation of the airflow rate through a cleaning unit. While real world studies of its application are limited \citep[e.g.][]{McLean1961}, laboratory and modelling studies suggest that, provided sufficient efficacy of UV in deactivating the virus, the method can be effective at mitigating airborne transmission \citep[e.g.][]{Escombe2009, Noakes2015}. Upper room UV is likely to be at its most effective in poorly ventilated spaces \citep{Noakes2015} and can be installed reasonably easily at a low cost. However sizing a system is not always straightforward, and it is essential that systems are installed by professionals who also ensure that occupied zone UV-C irradiation levels are safe. 

The method depends on the virus reaching the upper section of the room which is exposed to UV light. Therefore, appropriate internal flow patterns are required  in order to ensure that the virus is transported through the unit and deactivated. 
The process is complicated by the large range in particle sizes (see \S\ref{sec:dropaero}) produced by an infected person, where larger particles require higher vertical velocities to be carried to the upper section. Therefore, while a large range of particle sizes may well be deactivated, there is a danger that a certain fraction remains. Studies such as \citet{Nardell2008} have demonstrated that the method can be implemented with minimal risk to occupants due to UV exposure. Further, ``ozone-free'' UV lamps can be used to avoid adverse effects on indoor air quality due to the production of ozone. 
Only real world trials will prove the efficacy of the method 
(see \S\ref{app:uvgi}) 
and costs (of purchase, installation and maintenance) should always be compared to the equivalent costs of appropriate upgrades to the ventilation provision to the indoor space.


\subsection{Summary}

Each of the engineering control measures considered here have their advantages and limitations, and none are able to entirely eradicate the risk of transmission. Upper room UV holds promise as a method to significantly reduce the risk of transmission, particularly in poorly ventilated spaces. However further effort is required to demonstrate its effectiveness against SARS-CoV-2 in real world scenarios. Air ionisation is an emerging technology, but there are unresolved concerns regarding its impact on indoor air quality, primarily due to the production of ozone, and there is little evidence that it is effective against viruses. The addition of air filters that are effective against viral transmission to the existing ventilation system are likely to require complementary modifications to the ventilation system to, at the very least, account for the changes in pressure drops across the system. Independent filter units can be used, however in order to maximise their impact in larger spaces an understanding of the airflow patterns within the space is useful.  Filters are an established technology which are likely to reduce risk without any detrimental impact to occupants if appropriate maintenance can be ascertained and achieved. 
Using desk fans to increase the mixing of air within a room is easily applied and may be a very affordable measure, however their merits are questionable in the absence of evidence of stagnant zones. Ceiling fans provide the benefit of increased vertical mixing within the room. However in the case of both desk and ceiling fans there are uncertainties as to their impact on re-suspension and deposition of virus-laden particles and the associated impact on transmission risk. The use of screens can be effective to mitigate spread via the droplet route, but their use is problematic as a measure against airborne transmission since impacts on the circulation of air and the local ventilation provision is unpredictable without detailed bespoke study. Moreover, a full assessment of a screen's impact on fire safety would be required ahead of any installation.

The primary control measure for airborne transmission indoors is ventilation and achieving adequate outdoor air supply rates to ensure an acceptable level of risk should be the priority. However, where additional risk reduction is required or desired other measures can be implemented to help mitigate the spread of COVID-19. Where the intention of these measures is to reduce the risk of airborne transmission their costs should always be compared to the cost of upgrading the existing ventilation provision. Moreover, these measures should always be implemented in addition to any existing measures to reduce the spread of infection via all routes of transmission, e.g. adequate hand hygiene \& cleaning, the wearing of face coverings or other personal protection equipment (PPE) and social distancing.


\section{Author contributions}

{This document results from the work and discussions led by Professors Paul Linden (pfl4@cam.ac.uk) and Christopher Pain (c.pain@imperial.ac.uk) under Task 7 (Environmental and aerosols transmission) within the Royal Society's `Rapid Modelling of the Pandemic project' (RAMP). 

Dr Henry Burridge (h.burridge@imperial.ac.uk) led the compilation and editing of this document. Drs Megan Davies Wykes, Hywel Davies \& Shaun Fitzgerald, and Professor Paul Linden each acted in the capacity of supporting editors; all members of RAMP Task 7 contributed directly or indirectly to the production and editing of this body of work. Further notable contributions include:}

Production of executive summary:
\begin{itemize}
    \item Henry Burridge (h.burridge@imperial.ac.uk)
\end{itemize}
Production of text concerning COVID-19 transmission indoors and transmission via droplets and aerosols:
\begin{itemize}
    \item Rajesh K. Bhagat (rkb29@cam.ac.uk)
    \item Marc Stettler (m.stettler@imperial.ac.uk)                \item Prashant Kumar (p.kumar@surrey.ac.uk)
\end{itemize}
Production of text concerning transmission via surface contacts \& fomites, and the appendix `SARS-CoV-2 on hard surfaces':
\begin{itemize}
    \item Ishanki~De~Mel (i.demel@surrey.ac.uk)
    \item Panagiotis~Demis (p.demis@surrey.ac.uk)
    \item Allen~Hart (ah792@bath.ac.uk)
    \item Yyanis~Johnson-Llambias (yjl31@bath.ac.uk)
    \item Marco-Felipe~King (M.F.King@leeds.ac.uk)
    \item Oleksiy~Klymenko (o.klymenko@surrey.ac.uk)
    \item Alison~McMillan (a.mcmillan@glyndwr.ac.uk)
    \item Piotr~Morawiecki (pwm27@bath.ac.uk)
    \item Thomas~Pennington (tp373@bath.ac.uk)
    \item Michael~Short (m.short@surrey.ac.uk)
    \item David~Sykes (david67sykes@gmail.com)
    \item Philippe~Trinh (p.trinh@bath.ac.uk)
    \item Stephen~Wilson (s.k.wilson@strath.ac.uk)
    \item Clint~Wong (clint.wong@maths.ox.ac.uk)
    \item Hayley~Wragg (hw454@bath.ac.uk)
\end{itemize}
Production of text concerning social distancing indoors:
\begin{itemize}
    \item Megan Davies Wykes (msd38@cam.ac.uk)
\end{itemize}
Production of text concerning guidance on face masks and coverings, and the appendix `Summary of current recommendations for the use of face masks or coverings':
\begin{itemize}
    \item David Sykes (david67sykes@gmail.com)
    \item Chris Iddon (naturalventilation@cibse.org)
\end{itemize}
Production of text concerning source reduction through timetabling, and the purging between events:
\begin{itemize}
    \item Henry Burridge (h.burridge@imperial.ac.uk)
\end{itemize}
Production of text occupancy behaviour on indoor air movement:
\begin{itemize}
    \item Andy Woods (aww1@cam.ac.uk)
    \item Nicola Mingotti (nm441@cam.ac.uk)
    \item Neeraja Bhamidipati (nb522@cam.ac.uk)
\end{itemize}
Production of text concerning ventilation and the airborne transmission route:
\begin{itemize}
    \item Henry Burridge (h.burridge@imperial.ac.uk)
\end{itemize}
Production of text concerning local air movement and purification:
\begin{itemize}
    \item Huw Woodward (huw.woodward@imperial.ac.uk)
\end{itemize}
Production of text concerning upper room Ultra-Violet Germicidal Irradiation:
\begin{itemize}
    \item Huw Woodward (huw.woodward@imperial.ac.uk)
\end{itemize}
Production of the appendix `Natural inputs to ventilation provision':
\begin{itemize}
    \item Chris Iddon (naturalventilation@cibse.org)
\end{itemize}
Production of the appendix `UVGI and COVID-19':
\begin{itemize}
    \item Clive Beggs (C.Beggs@leedsbeckett.ac.uk)
\end{itemize}
Production of the appendix `Aerosols in the context of singing, and woodwind \& brass musical instruments':
\begin{itemize}
    \item Alison~McMillan (a.mcmillan@glyndwr.ac.uk)
\end{itemize}


\appendix

\section{Natural inputs to ventilation provision} \label{app:NV}

By design well-sealed buildings have ventilation provision that has been specified and installed, as per our guidance (\S\ref{sec:sum} and \S\ref{sec:VentGuide}) we recommend this provision is reviewed in light of COVID-19 prior to winter 2020. Many more buildings are, by design or otherwise, not well-sealed and in such cases the increased supply of outdoor air by natural means (e.g. through opening windows, doors, vents, louvres, etc…) may offer potential to mitigate risk of airborne transmission. However, these naturally driven flows rely on driving winds (which in winter are typically colder than comfortable indoor temperatures) or buoyancy (arising from temperature differences between indoor and outdoor aid) and so if not managed carefully will lead to either significant increases in heating bills and/or zones where cold draughts are uncomfortable. In this appendix, we seek to discuss practical means by which these natural flows can be exploited to mitigate risk but in a manner which attempts to balance the potentially negative consequences for indoor experience and energy consumption.


\subsection{Natural Ventilation in winter} \label{app:NV:Winter}

The amount of outdoor air that can be reasonably provided during winter may be less than in the summer due to impacts on indoor air temperature and occupant comfort. As there is a high confidence that, compared to adequately ventilated spaces, poorly ventilated spaces increase the risk of SARS-CoV-2 transmission via the far field ($>$2m) airborne route, it is important to ensure that poorly ventilated spaces are avoided. 

Approved Document Part F (ADF) sets out what, in ordinary circumstances, may be accepted as reasonable provision for compliance to the Building regulations: \textit{There shall be adequate means of ventilation provided for people in a building} \citep{ADF}. At the least, it is important to ensure that adequate ventilation is provided year round \citep[poor indoor air quality also negatively impacts health, wellbeing and productivity][]{BRE2019}. Wherever possible outdoor air flow rates should be maximised (i.e. increased more than the ADF provision rates for adequate airflow) where it is reasonable to do so. 
For low occupancy indoor spaces it may be that an airflow rate measured in litres per second per person (l/s/p) only provides a low overall flow rate and in these cases a minimum flow rate should be considered \citep{Jones2020}.

However increased ventilation in the winter may lead to unwanted occupant behavioural responses that result in ventilation provision being deactivated or minimised. For example, increased ventilation could result in colder indoor environments or cold draughts resulting in occupants closing or turning off ventilation provision. Thus, the goal of increasing ventilation provision results in no, or little, ventilation provision if such behavioural responses are elicited. Ventilation plays a key role in dilution of airborne SARS-CoV-2 encapsulated in respiratory aerosol --- poorly ventilated spaces increase the risk, although there is a diminishing law of returns in increasing the ventilation rate. The potential benefits of increasing ventilation to a poorly ventilated space are greater than increasing a well-ventilated space by the same amount \citep{Jones2020}.

In winter, the driving forces for natural ventilation are usually greater \citep[pressure differences caused by wind and indoor/outdoor temperature differences][]{CIBSEA,RN2} and therefore, to deliver the same flow rate, openings do not need to be as wide in the winter as they would need to be in the summer. What follows provides guidance for modulating various ventilation openings to deliver outdoor airflows whilst minimising occupant discomfort.

\subsubsection{Adjusting airflow through natural ventilation openings} \label{app:NV:Winter:01}
\textit{A single set of high- and low-level openings:}
In this configuration it is preferable to open the high level vents first to provide outdoor air, and to open the low level windows to further maximise airflow when reasonable. The turbulent plume of cooler outdoor air entering through high level vents will entrain warm room air as it falls under gravity, tempering the air before it enters the occupied zone \citep{turner-1973}. A helpful draught plume calculator is available in the BB101 calculation tools, which enables this effect to be measured \citep{BB101Tool}. A safe means of opening and closing high level vents should be supplied in workplaces \citep[][, regulation 15]{Workplace1992}.

\textit{Multiple openable windows and/or vents:}
Where a room has multiple openable vents, it may be possible to deliver the adequate ventilation provision through opening of just one vent. However, it is usually possible to create a more comfortable indoor environment, with respect to draughts, if the airflow is achieved through opening all the vents by a smaller amount than that necessary in the scenario of a single set of openings as described above.
If there are openable vents at both high and low level, then the principle of opening as many high level vents should initially be considered (see above).

\textit{Sash windows:}
As with high and low level windows, it is better to open the high level sash to provide openings at the top of the vent to encourage entrainment of outdoor air with the warm indoor air in the first instance. To further increase outdoor airflow the bottom sash can also be opened.

\textit{Other vents:}
In addition to windows, there are other vents and louvre systems that can be modulated and similar to windows, the principles of opening high level vents and multiple vents a small amount should be considered in the first instance.

\textit{\co sensors:}
High levels of \co (i.e. values greater than 1\, 500\,ppm) are indicative of a poorly ventilated space, and therefore \co can be a useful indicator of a space that is lacking adequate ventilation. However, low \co concentrations are not necessarily indicative of a well ventilated space. 
If \co sensors are to be deployed they should be Non-dispersive Infra-red (NDIR) \co sensors, which actually detect \co in the space, rather than the less expensive e\co sensors that do not detect \co and infer a \co concentration by measuring room volatile organic compound (VOC) concentrations instead.

\subsubsection{Occupant Comfort} \label{app:NV:Winter:02}
A person’s sensation of warmth is influenced by the following main physical parameters, which constitute the thermal environment:
\begin{itemize}
    \item air temperature,
    \item mean radiant temperature,
    \item relative air speed, and
    \item humidity.
\end{itemize}
Besides these environmental factors there are personal factors that affect thermal comfort:
\begin{itemize}
    \item metabolic heat production, and
    \item clothing.
\end{itemize}
It is possible to adjust the personal environmental factors to improve occupant comfort, particularly where outdoor air supply may decrease occupant comfort. CIBSE Guide A describes the predictive mean vote (PMV) method of measuring occupant thermal comfort, which can be used to see how changes to apparel and metabolic activity can improve occupant thermal comfort \citep{CIBSEA}.

\textit{Clothing:}
Clothing provides a layer of insulation that contributes to the occupant’s perception of comfort and is dependent upon the material and fit. Typically winter wear would have a value of 0.8 to 1.0\,clo, although studies in recent decades in Europe and the UK have found values generally at the lower end. This may be a result of occupants acclimatising to, and expecting, warmer indoor environments in the winter. For example compare typical winter office clothing of the 21$^{\textrm{st}}$ Century, with that in vogue at the turn of the 20$^{\textrm{th}}$ Century. To improve occupant comfort, particularly in naturally ventilated indoor spaces, occupants should be encouraged to dress appropriately and relaxation of dress codes should be considered if necessary to allow warmer clothes to be worn.


\textit{Metabolic rate:}
Sedentary activities have a relatively low metabolic activity which can contribute to occupant thermal discomfort. Encouraging periods of activity, to move around the room, or partake in some light exercise, will help to improve thermal comfort – as well as aiding in meeting display screen equipment (DSE) regulations which require regular breaks or changes in activity when using DSE \citep{DSE1992} and recommendations for regular movement to improve health.

\textit{Position of occupants in relation to openable vents:}
Where it is possible, increasing the distance of occupants from openable vents provides more time for incoming cool air plumes to entrain with warm room air prior to entering the occupied zone. 


\subsection{Summary}

We summarise by first suggesting that when attempting to reduce the risk of airborne spread of COVID-19 by increasing ventilation through natural means (e.g. opening windows wider) occupants comfort be given due consideration; otherwise, the chances that occupants act to intervene and unintentionally increase infection risk (e.g. by closing windows) is significant. To that end, the principles that high-level ventilation openings should be preferentially used, and that multiple smaller openings are preferable to a lesser number of larger openings, should be employed. Given that the maximum capacity of heating systems are broadly fixed and should be designed to accommodate the ventilation air required to deliver adequate indoor air quality, we do not encourage large increases in energy consumption, to ventilate an indoor space, beyond that which is absolutely necessary. 

\section{SARS-CoV-2 on hard surfaces} \label{app:surfaces}

\begin{figure}[htbp]
    \centering
    \includegraphics[scale = 0.4]{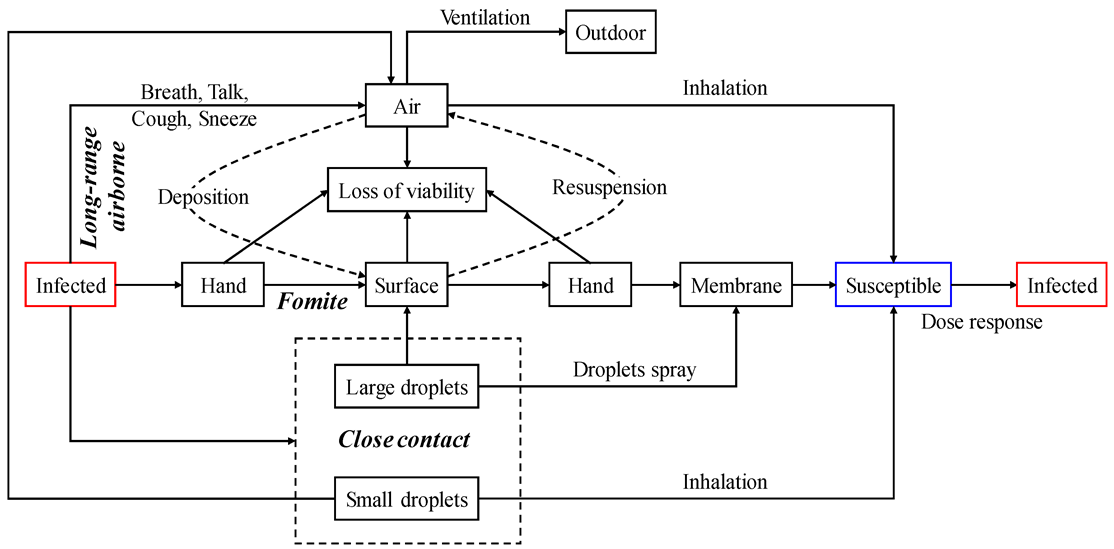}
    \caption{Three potential transmission pathways for respiratory viruses. [obtained from \citet{Zhang2018}]}
    \label{fig:pathways}
\end{figure}

\noindent Respiratory viruses can be transmitted from an infected individual to a susceptible host through three main pathways illustrated in Figure \ref{fig:pathways}. In this report, we focus on pathogen transmission mediated by surfaces of inanimate objects known as \emph{fomites} that, when contaminated with an infectious agent, can transfer disease to a new host. Fomite contamination can occur through: (i) direct contact with infected individuals and their bodily secretions/fluids; (ii) bioaerosols generated via breathing, talking, sneezing, or coughing; (iii) by an airborne virus that settles after a contaminated fomite is disturbed. A susceptible individual coming into contact with a contaminated fomite surface may pick up the pathogen and self-inoculate leading to infection.

Risk of viral infection through fomites can be assessed on the basis of three main considerations which determine the quantity of pathogen picked up upon touching the surface:
\begin{enumerate}{}
	\item The concentration of virus on the surface, which is affected by
    \begin{enumerate}{}
        \item source of virus (\emph{e.g.} sneezing, coughing or touching the surface by infected individuals)
        \item time elapsed since surface contamination
        \item surface characteristics (\emph{e.g.} material, smoothness, roughness, porosity) which affect the survival of the virus
        \item environmental and skin conditions (\emph{e.g.} temperature, humidity, perspiration)
        \item cleaning and/or disinfection activities since contamination
    \end{enumerate}
    \item The probability/frequency of contact with a contaminated fomite (\emph{e.g.} through touch by hand); probability of virus transfer from hands to the mucous membranes of mouth or nose
    \item The individual susceptibility to the virus (dose response curve)
\end{enumerate}
Many of the above factors are hard to quantify or control, and hence there is limited existing knowledge on transmission via fomites. In most settings the exact pathway of respiratory virus transmission is not known because of continuous interplay between the surface and droplet/airborne transmission pathways through droplet deposition and possible re-suspension of the pathogen. Although most of the evidence of fomite transmission is indirect, and no evidence has yet been reported for SARS-CoV-2, it is generally accepted that viruses can survive on hard surfaces for prolonged periods of time, that they can be transferred between fomites and hands, and that hands come into contact with mucous membranes, thus enabling this transmission pathway \citep{Boone2007, Kutter2018}.

The report will not cover aspects related to aerosol deposition of the virus (including the effects of ventilation on the rate thereof) which, although crucial for the quantification of risk of infection, is not within the remit of Subgroup 4. The topic of virus re-suspension will also not be covered due to lack of evidence in the literature.

\subsection{Surface deposition and contact frequency}

\begin{figure}[htbp]
    \centering
    \includegraphics[scale = 0.5]{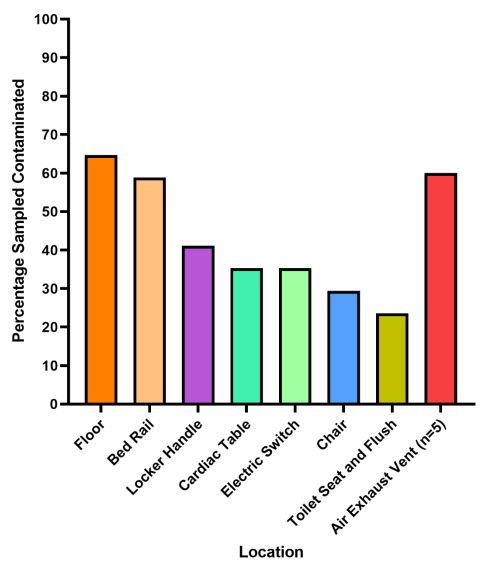}
    \caption{Percentage of contaminated swabs from surface samples in a hospital room housing a SARS-CoV-2 patient. [obtained from \citet{Chia2020}]}
    \label{fig:loc}
\end{figure}

\noindent The risk of infection through fomites depends on the frequency of contact with contaminated objects. It is therefore important to have quantitative information about the level of contamination of different surfaces as well as the likelihood of susceptible individuals coming into contact with them. Figure \ref{fig:loc}, from \citet{Chia2020} shows that SARS-CoV-2 is easily transported and deposited onto surfaces in a hospital setting (airborne infection isolation rooms), with certain surfaces that are commonly handled showing traces of the virus. Note that the figure indicates the likelihood of finding the virus on different surfaces but does not reflect actual viral loads found on them.

\citet{Zhang2018} have reported a large study of surface touch in a student office, in which they collected over 120,000 touch events. They introduce a classification of surfaces into high-touch (\emph{e.g.} those touched multiple times by the same person) and high-risk surfaces (touched by multiple individuals) as well as private (\emph{e.g.}, a mobile phone) and public (\emph{e.g.}, door knob, light switch, etc.). The study revealed that 96.7\% of touches in an office setting involved private surfaces and only 1.2\% of the touches were of public surfaces which play the most important role in fomite transmission. The authors use a network approach to assess `distances' between surfaces in the surface touch network and the dynamics of spread throughout the office. The paper also reports frequencies and average durations of contact with various surfaces in a student office which provides rich data for modelling transmission.


\subsection{Factors affecting survival of virus on surfaces}
\label{A2}
The survival of viruses on surfaces is strongly impacted by the type of surface and surface material upon which it is deposited as well as by a number of environmental factors such as ambient temperature, light, humidity, and pH (if in droplet form on surface). Additionally, factors such as type of contamination (fingertip, droplet, etc.) and the concentration of the virus inocula (relevant to droplets) are important \citep{Warnes2015}. For SARS-CoV-2, the virus was found to be active across a wide range of pH, from 3 to 10 \citep{Chin2020} and so changing the pH to deactivate the virus may not be a feasible option.

\subsubsection{Surface material}

\begin{figure}[t]
    \centering
    \includegraphics[width=1.0\textwidth]{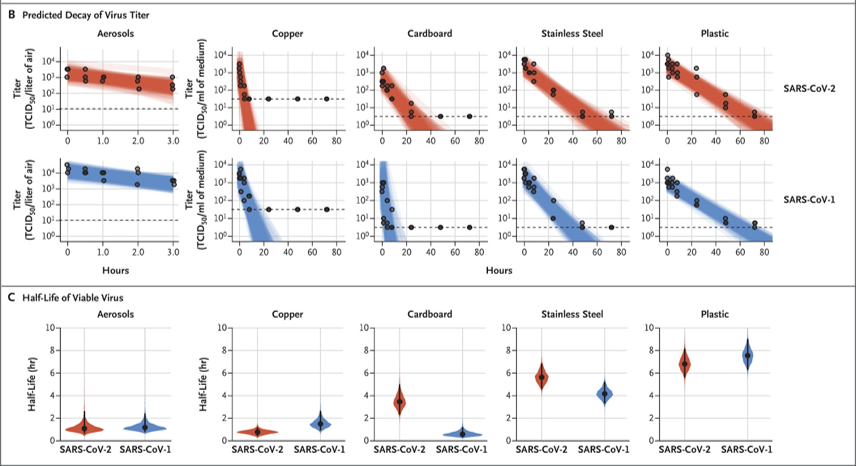}
    \caption{ Exponential decay and half-lives of SARS-CoV-2 virus (red) and SARS-CoV-1 virus (blue) on different surfaces. [obtained from \citet{VanDoremalen2020}]}
    \label{fig:mesh1}
\end{figure}

The type of surface can affect the stability of the SARS-CoV-2 virus. A study has investigated the stability of both SARS-CoV-1 and SARS-CoV-2 viruses on plastic, stainless steel, copper and cardboard surfaces under conditions of 21--23$^{\circ}$C and 40\% relative humidity \citep{VanDoremalen2020}. Note that the initial concentration of the virus used on the surface is $10^{3.7}$ TCID$_{50}$ per ml of medium. The virus decayed exponentially with time on each of these surfaces, as seen in Figure 1. The half-lives of the virus on plastic and stainless steel have been reported as 6.8 hours and 4.6 hours, respectively. From Figure \ref{fig:mesh1}, the half-lives of the virus on cardboard and copper are approximately 4 hours and 1 hour, respectively. This study also reports that the amount of infectious virus greatly reduced within 72 hours and 48 hours for plastic and stainless steel, respectively, and that no infectious virus was found on cardboard and copper after 24 hours and 4 hours, respectively.

\begin{table}[!ht]
    \centering
    \caption{Time taken (approximate) for no infectious virus to be found on each surface \citep{Chin2020}. \label{tab:surfaces}}
    \begin{tabular}{r l}    
     \emph{Surface} & \emph{Time taken (hours)}\\
        Stainless Steel & 168\\
        Plastic & 168\\
        Wood & 48\\
        Cloth & 48\\
        Paper & 3\\
        Tissue & 3\\
        Glass & 96\\
        Bank note & 96\\
        Mask (inner layer) & 168\\
        Mask (outer layer) & $>$168\\    
    \end{tabular}    
\end{table}

The stability of the SARS-CoV-2 virus on stainless steel, plastic, wood, cloth, paper, tissue, banknote and inner and outer layers of a surgical mask has been evaluated by \citet{Chin2020}. Each surface was pipetted with a virus culture of approximately $10^{7.8}$ TCID$_{50}$ per ml of medium (higher than the concentration used in \citet{VanDoremalen2020}), maintaining conditions of 22 $^{\circ}$C and 65\% relative humidity. This study has not reported the half-lives of the virus on each of these surfaces; however, the authors have reported the time it took for no infectious virus to be detected on each surface. Table 1 summarises their findings.

Results also show that both inner and outer layers of the mask retain infectious virus for a significantly long time \citep{Chin2020}. However, it is possible that their experimental approach does not truly reflect a real scenario, as the virus culture has simply been pipetted onto the inner layer of the mask. It is not clear from this study whether it is possible for the virus medium (especially in droplet form) to seep through from the outer layer to the inner layer and retain a high concentration of the virus within the inner layer. Furthermore, it is unlikely that the inner layer of the mask would remain at 22 $^{\circ}$C in reality, as exhaled air from the nose would increase the temperature.

The study also concludes that the virus remains stable for longer on smooth surfaces \citep{Chin2020}; however, note that roughness/smoothness has not been quantified. Exceptions to this have been seen in a study performed on the Coronavirus 229E strain \citep{Warnes2015}. The authors of this paper report that an increase in the copper content of a surface (such as in brass) can significantly reduce the time taken for the virus to be inactivated. They claim that this inactivation is due to the release of copper ions and reactive oxygen species. It is possible that this phenomenon is the cause of the low half-life reported for SARS-CoV-2 on copper, as seen in \citet{VanDoremalen2020}. 

\citet{Biryukov2020} has studied the half-life of SARS-CoV-2 on stainless steel, plastic, and nitrile gloves (all nonporous surfaces) under various temperatures and relative humidities (see section on Effects of Temperature and Humidity). Interestingly, their results do not show significant variations in viral decay rates on these surfaces when environmental conditions are kept the same, and conclude that surface type does not impact viral stability. It is unknown why these results are different from those of \citet{VanDoremalen2020} and \citet{Chin2020}, highlighting the need for further experiments in this area.

\subsubsection{Effects of temperature and humidity}

Experiments on SARS-CoV-2 under varying temperature have showed that the virus is less stable under increasing temperature \citep{Chin2020,Dietz2020}. \citet{Chin2020} show that the inactivation of the virus can be reduced to 5 minutes at a temperature of 70 $^{\circ}$C, whilst remaining stable for over 14 days at 4$^{\circ}$C. 

\citet{Biryukov2020} have analysed the effects of different temperatures (24$^{\circ}$C and 35$^{\circ}$C) and relative humidities (20\%, 40\%, and 60\%) on SARS-CoV-2, diluted in saliva and deposited on surfaces in droplet form. They confirm that the half-life of the virus is reduced when temperature and humidity are increased, either independently or in combination (the lowest half-life of approximately 2 hours has been reported for the temperature-humidity combination of 35$^{\circ}$C and 60\%). While they also analyse the effects of different droplet sizes and deposition on three different nonporous surfaces (stainless steel, plastic and a nitrile glove), they conclude that the size of the droplet and surface type do not significantly impact the half-life.

Studies on other strains of coronaviruses show similar trends \citep{Kampf2020, Ren2020}. One study reports that temperatures of 30 – 40$^{\circ}$C is sufficient to reduce time taken to inactivate viruses \citep{Kampf2020}. \citet{Casanova2010} also concludes that increasing both temperature and humidity have contributed to faster inactivation/decay of coronaviruses. A review on viruses on surfaces has reported that the viruses are more stable at relative humidity levels below 50\% \citep{Vasickova2010}. \citet{Dietz2020} predicts that higher relative humidity facilitates larger droplets containing viruses to settle on surfaces more quickly, thus reducing airborne transmission \citep{Dietz2020}.

Based on these results, we can conclude with a high level of certainty that virus stability on surfaces decreases with increasing temperature and humidity. In reality, relative humidity levels should be increased cautiously as levels of 80 \% or greater can lead to mould growth \citep{Dietz2020}. 

\subsubsection{Effects of surface finish, texture or roughness}

All surfaces are inherently rough at some length-scale, and the particular quality 
of the surface is determined by the method of manufacture and any finishing processes, 
such as polishing. Even with the most sophisticated finishing processes, it has to 
be accepted that some roughness remains, and that the length-scale of the roughness 
features might be of a similar size to that of some droplets falling onto that surface. 
This then prompts questions as to differences in evaporation, adhesion, surface tension, 
and touch transferability of viral material contained in droplets that land proud on a 
smooth (or relatively smooth) surface, to those that land or flow down into roughness 
grooves in the surface.

\begin{figure}[t]
    \centering
    \includegraphics[width=0.8\textwidth]{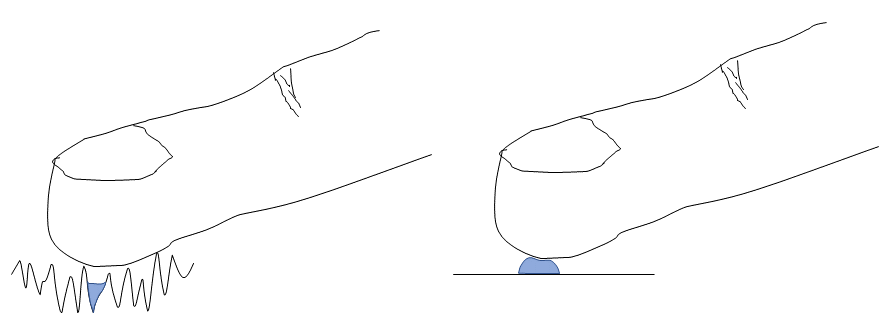}
    \caption{ Illustration of how a droplet in a surface roughness groove 
(left) can be inaccessible to finger touch compared with a droplet on a smooth 
surface (right). }
    \label{fig:roughness1}
\end{figure}

The science of surface characterisation, rubbing friction and surface lubrication is 
called ``Tribology'', and was established by \citet{bowden1950}. Surface texture 
(deliberate or arising from roughness) is characterised in a number of ways, but 
a common metric is $R_a$, which is defined as the arithmetical mean deviation of 
the surface profile being assessed. Other metrics consider peak to trough distance, 
root-mean-square, statistical aspects (skew and kutosis), or reflect more directly the 
method of measurement (sizes and numbers of peaks and troughs over a given measurement 
area). Another approach recognises a fractal-like distribution in roughness and 
characterises roughness based on Box-Fractal dimension 
\citep{mandelbrot1984,mandelbrot2006}. For the present purposes, a suitable 
characterisation should express the fraction of surface area which would be likely 
to catch droplets of a given size within a trough-like region. On the basis of 
simple geometric considerations, one could imagine that for some combination of 
surface roughness types and droplet size and number, the contamination would mainly 
fall beneath the level at which a finger could press into the surface to touch, as 
illustrated in Figure \ref{fig:roughness1}. 

Another aspect that would influence the survival of infected droplets on surfaces 
is the adhesion of the droplet to the surface. The authoritative text on the 
subject of adhesion and adhesives is that by \citet{kinloch1987}. The geometric 
interlocking of interfaces is not thought to be a strong factor in surface adhesion, 
but in considering the removal of droplets or dried droplet residue from a surface, 
there is a question to be asked of whether such material can be wiped off cleanly, 
or whether some material is forced more deeply into grooves on the surface. 
Understanding adhesion mechanisms at the micro-scale could provide some valuable 
insight into the mechanical aspects of the cleaning process.

\subsection{Modeling approaches}

\subsubsection{A model to predict viral transfer from surfaces}

Transfer of microorganisms to hands from contaminated surfaces can be modelled mechanistically \citet{King2020}:

\begin{equation}
C_H = \lambda_{S\rightarrow H} C_S
\end{equation}

where $C_H$ is the concentration of virus on fingers, $\lambda_{S\rightarrow H}$ is the percentage transfer during a single contact which is pathogen and surface dependent and $C_S$ is the concentration on the surface. This can be extended to multiple contacts:

\begin{equation}
    C_{n}= C_{n-1} + \lambda_{S\rightarrow H} (C_S-C_{n-1})
\end{equation}

where $n$ is the $n^{\textrm{th}}$ surface contact. $\lambda_{S\rightarrow H}$ is difficult to measure accurately and while experiments are underway to quantify this for SARS-CoV-2, surrogates can be used. \cite{Kraay2018} suggest $\lambda_{S\rightarrow H}$ for influenza may lie between 4\% and 16\%, whilst rhinovirus may range up to 40\%. Data of finger to mucosa transfer is much more scarce but \cite{Rusin2002} estimates a viral surrogate at 33.9\% (unpublished standard deviation was later found to be 16\%).

\subsubsection{A model to predict viral decay rate}
A simple thermodynamic model has been developed by \citet{Yap2020} which reasonably predicts the decay rate of different coronaviruses including SARS-CoV-2 at different temperatures. By assuming an exponential decay for the quantity of pathogens, $C$, with 
\[
C = C_0e^{-kt},
\]
where $C_0$ is the initial quantity, $t$ is time (seconds) and $k$ is the decay rate. The decay rate is assumed to satisfy a relation (the \emph{Arrhenius equation}) that is typical for the thermal denaturation of protein, namely 
\[
k = Ae^{-{E_a}/RT},
\]
where $T$ is the ambient temperature (Kelvin), $R$ is the Gas constant, $E_a$ is the activation energy ($\mbox{J}/\mbox{mol}$) and $A$ is an arbitrary frequency in the limit of high temperature. In particular, they found a universal relation (a \emph{Meyer-Neldel relation}) between $E_a$ and $A$ for different coronaviruses, namely
\[
\log(A)= 0.394E_a-5.63,
\]
based on experimental data from previous work. Although it has been assumed that this relation holds for all temperatures, the authors claimed that this is a reasonable assumption for proteins and, most importantly, this relation also has good agreement with previous findings. For SARS-CoV-2, we can readily predict the decay rate using the experimentally fitted value $E_a=135,692$ $\mbox{J}/\mbox{mol}$. 

The authors have also suggested that this model can be extended to incorporate other catalytic effects \emph{e.g.} humidity and surface properties by establishing further relations with the activation energy.

\subsubsection{Droplet evaporation from a surface}
 Numerous studies exist for the size of saliva droplet upon expulsion from the mouth (via breathing and coughing), with results between studies showing considerable variance. However, by assuming the mechanical properties of pure water, the mathematics is well-established for predicting the time scale for such droplets to evaporate, namely 
\[
t_{\mathrm{evap}}\sim\frac{\rho\theta}{D(1-H)c_{\mathrm{sat}}}R_0^2,
\]
where $\rho$ is the density of water, $\theta$ is the contact angle, $D$ is the diffusion coefficient, $H$ is the relative humidity, $c_{\mathrm{sat}}$ is the saturation concentration of vapour in air and $R_0$ is the initial droplet radius [see \citet{Dunn2009} and \citet{Bhardwaj2020} for further details on how these parameters depend on temperature and humidity]. 

There is evidence of a slower drying time of saliva as compared to pure water; In \citet{liu2017evaporation}, saliva droplets released onto a Teflon-printed slide showed a $20\%$ longer evaporation time than those consisting of pure water. These results were especially pronounced at lower humidities and over longer time periods. Additionally, the same study notes evidence that droplets produced by coughing may be more likely to contain a higher solute concentration, due to the presence of pulmonary mucus. Droplets produced by coughing are also larger in size \citep{Chao2009}, indicating a greater likelihood to be deposited on surfaces.
\par
Under the assumption of pure water properties, such physical modeling indicates total droplet evaporation within seconds, which is multiple orders of magnitude less than the typical lifetime of SARS-CoV-2 on various surfaces. The results suggest that SARS-CoV-2 do not require a saliva droplet to survive although we have not found any evidence corroborating this. However, this time scale can be of interest for frequently touched surfaces, \emph{e.g.} door handles. In such settings, a droplet with a high viral load can transfer from surface to skin by contact.

\subsubsection{Modelling of droplet residue and how its mechanical performance and 
attachment to surfaces could influence its survival}

At the present time, there has been no micro-mechanical characterisation of the 
material properties of partially-evaporated saliva droplets; however, if such 
properties were known then computational models of droplets undergoing evaporation 
could be made. At some point during this process, the droplet would first show higher 
viscosity, then become increasingly non-Newtonian as it transitions from being 
fluid-like to becoming a gel, or solid-like. The morphology of the droplet during 
that process would be interesting to study, particularly in regard to how the 
evaporation process drives the droplet shrinkage. 

It is fairly obvious that a rough surface provides more surface area than a smooth 
one, and so a fluid material spread over a rough surface would have more contact, and 
therefore be more greatly affected by heat conduction between the surface and the 
fluid. Depending on the surface tension characteristics, and how well the fluid 
wetted the surface, then there could also be greater fluid spread, and more area 
available for evaporation.

\begin{figure}[t]
    \centering
    \includegraphics[width=0.75\textwidth]{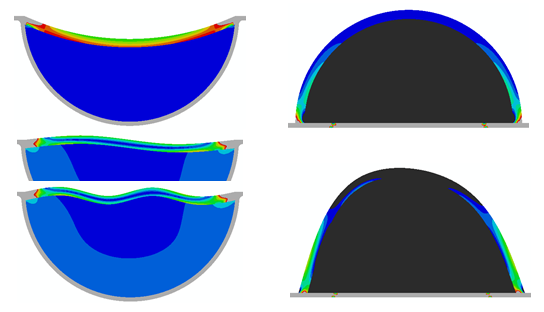}
    \caption{ Illustration of skin stresses in shrinking droplets: for droplets in 
surface roughness grooves (left) and droplets on a smooth surface (right). Notice that 
skin undulation or asymmetry potentially mitigates against the surface tearing mechanism.}
    \label{fig:roughness2}
\end{figure}

The following is mere supposition, but illustrates how having a better understanding 
the material properties would yield a better understanding of viral survival and the 
effect of surface roughness in that survival. Let us suppose that an outer layer of 
the droplet forms a skin or crust: an outer layer, exposed to the air, and therefore 
more dried out by the evaporation process. Let us suppose that one action of the 
skin is to protect or reduce evaporation of the inner material. Should that skin 
rupture, then the inner material would become exposed, and that would then form a 
skin, and the volume of inner material would be reduced. Encouraging the skin to 
rupture frequently could lead to faster evaporation and thus to complete drying out. 
Depending on the adhesion strength of the droplet to the walls of roughness grooves 
in the material, it is possible that the evaporation could lead to the skin becoming 
stretched and tearing. The sort of modelling result that such analysis could provide 
are given in Figure \ref{fig:roughness2}, but these are to be understood to be 
illustrative only, since the materials property data they are based on is fabricated. 
If the virus requires some moisture to survive then devising ways of obtaining greater 
degrees of drying would be beneficial to us. 

Alternatively, the more important aspect might be the ease of which a part of the 
droplet can be adhered to a finger and plucked from the surface. What are the 
relative adhesion strengths of the droplet to hard surface, and droplet to finger?  
Does finger dampness play a role in this?  When a droplet is picked up on a finger in 
this way, is it transferred whole, or does a part remain stuck to the surface?  If the 
latter is the case, then presumably the moister inner material is then exposed, and 
perhaps the virus is more readily transferred from the finger to other parts of the 
body, such as hand-to-face contact.

Given the potential value that credible geometrical, fluid and mechanical modelling 
could provide, micro-scale materials characterisation would enable a much greater 
understanding of the physical processes within the droplet, how the droplet is transferred, 
and how this influences the survivability of the virus within the droplet.

\subsection{Self-inoculation via hand-to-face contacts}
For an individual to get infected after contact with a contaminated surface, self-inoculation with a sufficient dose of the virus must occur. Typical hand-to-nose contact frequency occur about $2.5 \pm 2.2$/hour \citep{Wilson2020}. A study by \citet{Nicas2008} reports observed frequency of contact between individuals' hands and mucous membranes (eyes, lips and nostrils) and a model to quantify the risk of infection taking into account pathogen pick-up, inactivation on the hands with time and frequency of contact between hands and mucous membranes. The data reported are based on seasonal Influenza A virus.

\citet{Zhang20182} report a detailed study of transmission of Influenza A in a student office through person-to-person contact and surface touch behaviour. The authors consider airborne and fomite-mediated transmission pathways and report detailed data about surfaces and individuals involved including close contact (enabling transmission through large droplets) and surface touch observations. Based on a mathematical model describing both the aerosol and surface transmission pathways, the authors conclude that ~4.2\% of risk of infection by Influenza A is due to transmission via fomites. On the other hand, in the office 95.1\% of viruses were transmitted via private surfaces and only 4.9\% via public surfaces with desks being the most contaminated due to virus-laden droplets generated while talking, sneezing and coughing being deposited on desks (including computer keyboards, mice, etc. located on them).

It has recently been suggested that transmission through contact between a contaminated hand and the eye may not be the preferred way of infection by SARS-CoV-2 \citep{Liu2020}.

\subsection{Cleaning and disinfection}
\subsubsection{What is an effective detergent?}

Soap and water has been widely recommended to remove the SARS-CoV-2 virus from surfaces and skin. \citet{Chaudhary2020} investigates the mechanisms employed by the surfactants found in soap to disrupt the virus and conclude that soap and water is effective against the virus. However, \citet{Chin2020} implies that hand soap solution cannot inactivate the SARS-CoV-2 virus very well; that infectious virus can be found after 5 minutes of incubating in hand soap solution. However, they have not simulated hand washing and have simply added the virus culture to hand soap solution. The physical removal of the virus trapped in soap micelle while rinsing hands has thus not been simulated in this study. 

\subsubsection{What is an effective disinfectant?}
\citet{Kampf2020} reports that the following substances can readily inactivate the infectivity of coronaviruses (by 4log$_{10}$ or more):
Ethanol (78\%–95\%),
2-propanol (70\%–100\%),
2-propanol (45\%) with 1-propanol (30\%),
glutardialdehyde (0.5\%–2.5\%),
formaldehyde (0.7\%–1\%)
and povidone iodine (0.23\%–7.5\%).

Quaternary ammonium compounds (QAC) are known to be effective against both enveloped and non-enveloped viruses and as such are a common additive to disinfectants. Although they exhibit a residual biocidal effect over 2 to 4 hours after application which has been shown to be effective against norovirus, it is unknown whether enveloped coronaviruses are susceptible at these lower concentrations. Sodium hypochlorite requires a concentration of at least 0.2\%, whilst hydrogen peroxide requires a concentration of at least 0.5\% and left incubating for at least 1 minute. Chlorhexidine digluconate has been found to be ineffective against coronaviruses \citep{Kampf2020}. It has been highlighted that SARS-CoV-2 is an enveloped virus and therefore can be less resistant to disinfection than non-enveloped viruses \citep{Glasbey2020}. The United States Environmental Protection Agency has listed substances [see List N in \citet{USEPA2020}] that are potentially effective against SARS-CoV-2 \citep{Glasbey2020}. Neither the substances in this list nor the substances in \citet{Kampf2020} have been experimentally proven to be effective against SARS-CoV-2, although they have been proven for other strains of coronaviruses. Appropriate procedure should begin with cleaning with soap and warm water followed by application of a disinfectant in the above category.

\subsubsection{Light deactivation}
\citet{Dietz2020} reports that the impact of sunlight and UV light on the stability and decay of SARS-CoV-2 has not yet been investigated. UV-C light has been proven to inactivate other strains of coronaviruses within 10 minutes or less \citep{Bedell2016}. However, authors of this study state that UV-C light is best employed alongside regular cleaning of surfaces, and use of light cannot simply substitute effective cleaning practices. These measures also require thorough risk assessment and mitigation to ensure that occupants’ health is not compromised and exposure is minimised. 

\subsection{Science behind current UK government advice}
The UK government guidelines for cleaning of surfaces that are currently available do cite few scientific articles regarding fomite transmission and we can therefore conclude that this pathway is thought to be less significant than other transmission pathways.

The current government advice on cleaning in non-healthcare settings is to wipe hard surfaces with warm soapy water before applying detergent. The guidelines lack definitions for both substances and only suggest chlorine-based disinfectants. In this document, we have provided a list of potentially effective detergents and disinfectants (mostly alcohol-based) that have been verified by experiments on other strains of coronaviruses.

Waste that is suspected to be contaminated by COVID-19 should be stored for 72 hours before disposal. This quarantine period also applies to returned goods in shops and clothes that have been tried on. We assume that this duration is based on the results by \citet{VanDoremalen2020} which suggest that SARS-CoV-2 can remain active for 72 hours (see \S\ref{A2} on Factors affecting survival of virus on surfaces). We are unable to find any guidelines on parcels.

Working safely during coronavirus (COVID-19) (\href{https://www.gov.uk/guidance/working-safely-during-coronavirus-covid-19}{link}) contains sector-specific guidelines, including those for transport, schools and childcare settings. These sectors have all been asked to clean ‘regularly and frequently’---frequencies have not been quantified.

\subsection{Conclusions and key questions}
While there is still a lack of detailed and complete information regarding the behaviour of SARS-CoV-2 virus on surfaces, experimental studies have confirmed that certain environmental conditions (such as high temperature and humidity) can reduce the stability of the virus, thus reducing infectivity. Furthermore, some studies have shown which disinfectants are potentially more effective against the virus than others, what surfaces are typically infected, and for how long the virus remains on potentially infectious on surfaces.

There is a lack of clarity in the UK government's advice on what cleaning products and methods are effective against the virus and which surfaces should be cleaned more frequently. In order to accurately advise the public, as well as operators of key infrastructure, such as public transport and hospitals, it is important to combine these experimental results with models that can be used to inform decision-making. 
In order to build such models, information regarding the types of deposits on surfaces, the concentration of the virus in these deposits, which surfaces are handled more frequently, as well as dose response curves are vital.

\section{UVGI and COVID-19} \label{app:uvgi}

With the emergence of COVID-19 and the subsequent global pandemic there has been considerable interest in the use of UV light to disinfect blood plasma \citep{31Ragan,32Keil,33Eickmann}, medical equipment \citep{34Heimbuch,35Hamzavi,36Card,37Derraik} and air \citep{Morawska20}, in the hope that this might reduce transmission of the disease. In particular, UVGI, a long established technology which utilizes UV-C light at wavelengths close to 254\,nm \citep{39Reed}, appears to have considerable potential as an intervention for inactivating SARS-CoV-2 \citep{40Inagaki,Beggs2020,42Bianco} and other pathogenic coronaviruses \citep{33Eickmann,34Heimbuch,43Duan,44Darnell,45Darnell,46Kariwa,Bedell2016}. 

The biological impact of UV radiation is primarily due to the absorption of energetic photons by nucleic acids. DNA has an absorption spectrum which has a maximum in the 260 -- 265\,nm region, which rapidly declines thereafter as the wavelength increases \citep{48Harm}. At these wavelengths photons of light are absorbed by nucleic acids, both DNA and RNA, to form dimers (fused base pairs) \citep{49Beggs} that impair the replication of viruses, inhibiting their ability to cause infection. Dimers involving both pyrimidines (cytosine and thymine) and purines (adenine and guanine) are the principal photoproducts formed, with pyrimidine dimers, particularly those involving thymine, generally predominating \citep{50Jagger}. The formation of these dimers inhibits transcription of the viral genome, thus preventing synthesis of the necessary proteins required for the viral replication. The number of bases in the viral genome is an important factor for determining sensitivity to UV damage, with longer genomes presenting more target molecules, increasing the likelihood that the virus will be inactivated for a given level of UV exposure \citep{51Sagripanti}. As such, the SARS-CoV-2 virus, having a genome that is almost twice as long as the influenza viral genome, should in theory be more susceptible to UV damage than influenza. Having said this, because both are RNA viruses and do not have thymine in their genome, they will both tend to be less sensitive to UV damage compared to DNA viruses of similar genomic length. 

While most work that has been undertaken on UVGI has involved low-pressure UV lamps with a strong UV-C spectral emission at 253.7\,nm, in recent years devices that produce light at other wavelengths have also been investigated. Specifically, deep-UV (DUV) light emitting diodes (LEDs), which emit light at around 280\,nm on the boundary between UV-B and UV-C light, have been evaluated. With specific reference to COVID-19, UV light from DUV-LEDs has been shown to inactivate the SARS-CoV-2 virus in petri dishes \citep{40Inagaki}. Others have also demonstrated that UV-C light at 222\,nm can be used effectively to inactivate coronaviruses in aerosols \citep{52Buonanno}.\\
\\

\subsubsection{UVGI inactivation Kinetics}

At any point in time the amount of viral inactivation (disinfection) achieved for a given UV radiant flux (irradiance) can be described using the following first order decay equation \citep{53McDevitt}
\begin{equation}
    N_t = N_0 \times e^{-Z\,E\,t} \, ,    \label{eq:CB1}
\end{equation}
where \textit{N$_{0}$} and \textit{N$_{t}$} are the number of viable viral particles (virions) at time zero and \textit{t} seconds respectively; \textit{Z} is the UV susceptibility constant for the virus (m$^{2}$/J); \textit{E} is the irradiation flux (W/m$^{2}$); and \textit{t} is time in seconds. \\

The UV irradiation dose received by the virus is given simply by
\begin{equation}
H=E\times{}t  \, , \label{eq:CB2}    
\end{equation}
where \textit{H} is the observed UV irradiation dose (J/m$^{2}$).

By combining equations \ref{eq:CB1} and \ref{eq:CB2}, and rearranging we can obtain an equation for \textit{Z}, namely
\begin{equation}
Z=-\frac{1}{H}\times{}\ln\left(\frac{N_{t}}{N_{0}}\right)=-\frac{1}{H}\times{}\ln(f)  \, , \label{eq:CB3}    
\end{equation}
where:\hspace{15pt} \textit{f} is the survival fraction.

Because the relationship between the UV dose and the natural logarithm of the survival fraction is broadly linear for most viral species, it means that the behaviour of any given virus exposed to UV-C light can be succinctly described by the value of \textit{Z}, irrespective of the actual UV dose applied. As such, for any given viral species, if the value of \textit{Z} is known, then it should be possible to predict with reasonable accuracy how the virus will behave when exposed to a given UV-C dose in any context. Microbes that exhibit larger \textit{Z} values are more susceptible to UV damage, whereas those with small \textit{Z} values are more difficult to inactivate.

UV inactivation plots for most viral species tend to be straight lines, although some might exhibit a curve \citep{46Kariwa}. Notwithstanding this, the model described in equation \ref{eq:CB1} is still a good approximation for most viral species \citep{53McDevitt} up until the point where the ‘target’ becomes saturated with UV photons. At this point, because all the virions have already been inactivated, increasing the UV dose further has no effect and so the linear relationship between UV dose and the log reduction becomes decoupled, with the result that the \textit{Z} value no long applies.

Instead of quantifying UV inactivation in terms of survival fraction, many researchers, particularly those working in biology, describe the reduction in the microbial count in terms of log reduction, which can be converted to survival fraction as follows
\begin{equation}
f=\frac{1}{10^A}  \, , \label{eq:CB4}    
\end{equation}
where \textit{A} is the log$_{10}$ reduction in the number of virions.

\subsubsection{Susceptibility of SARS-CoV-2 to UVGI}

While a substantial amount of work has been undertaken on the UV irradiation of coronaviruses in various contexts \citep{40Inagaki,42Bianco,52Buonanno,54Walker}, relatively little work has been done specifically on the SARS-CoV-2 virus, and that which has been done has focused solely on the irradiation of the virus on surfaces \citep{55Signify}, in blood \citep{31Ragan,32Keil} or in liquids \citep{40Inagaki,42Bianco}. However, from the work that has been done to date, a clear and consistent picture emerges, namely that in comparison with SARS-CoV-1 and MERS-CoV, the SARS-CoV-2 virus appears to be relatively easy to inactivate with UV-C light \citep{Beggs2020}. This is clearly illustrated in table \ref{tab:CB1}, which shows the calculated \textit{Z} values for various UV-C (and deep-UV at 280\,nm) irradiation experiments involving SARS-CoV-1, MERS-CoV and SARS-CoV-2 suspended in liquids. From this, it can be seen that the adjusted mean \textit{Z} for SARS-CoV-1 was 0.00489 (SD = 0.00611) m$^{2}$/J, similar to that for MERS-CoV (\textit{Z} value = 0.00104 m$^{2}$/J), whereas by comparison SARS-CoV-2 appears much more susceptible to UV damage at 254\,nm (adjusted mean \textit{Z} = 0.14141 (SD = 0.09045) m$^{2}$/J) and 280\,nm (mean \textit{Z} = 0.03684\,m$^{2}$/J).\\

\begin{center}
\begin{table}
\begin{tabular}{|p{88pt}|p{60pt}|p{68pt}|p{68pt}|p{109pt}|}
\hline
\parbox{88pt}{\raggedright 
{\small Virus}
} & \parbox{60pt}{\raggedright 
{\small UV-C dose }

{\small (mJ/cm$^{2}$)}
} & \parbox{68pt}{\raggedright 
{\small Inactivation}

{\small (log reduction)}
} & \parbox{68pt}{\raggedright 
{\small UV susceptibility constant , Z}

{\small (m$^{2}$/J)}
} & \parbox{109pt}{\raggedright 
{\small Reference}
} \\
\hline
\parbox{88pt}{\raggedright 
{\small SARS-CoV-1}
} & \parbox{60pt}{\raggedright 
{\small $>$81}
} & \parbox{68pt}{\raggedright 
{\small $>$ log 0.602}
} & \parbox{68pt}{\raggedright 
{\small 0.00171}
} & \parbox{109pt}{\raggedright 
{\small Duan et al. \citep{43Duan}}
} \\
\hline
\parbox{88pt}{\raggedright 
{\small SARS-CoV-1}
} & \parbox{60pt}{\raggedright 
{\small 241}
} & \parbox{68pt}{\raggedright 
{\small log 1.4*}
} & \parbox{68pt}{\raggedright 
{\small 0.00134*}
} & \parbox{109pt}{\raggedright 
{\small Darnell et al. \citep{44Darnell}}
} \\
\hline
\parbox{88pt}{\raggedright 
{\small SARS-CoV-1}
} & \parbox{60pt}{\raggedright 
{\small 1446}
} & \parbox{68pt}{\raggedright 
{\small log 4.5*}
} & \parbox{68pt}{\raggedright 
{\small 0.00072*}
} & \parbox{109pt}{\raggedright 
{\small Darnell et al. \citep{44Darnell}}
} \\
\hline
\parbox{88pt}{\raggedright 
{\small SARS-CoV-1}
} & \parbox{60pt}{\raggedright 
{\small 4819}
} & \parbox{68pt}{\raggedright 
{\small log 4.1*}
} & \parbox{68pt}{\raggedright 
{\small 0.00020*}
} & \parbox{109pt}{\raggedright 
{\small Darnell \& Taylor \citep{45Darnell}}
} \\
\hline
\parbox{88pt}{\raggedright 
{\small SARS-CoV-1}
} & \parbox{60pt}{\raggedright 
{\small 40}
} & \parbox{68pt}{\raggedright 
{\small log 3.2*}
} & \parbox{68pt}{\raggedright 
{\small 0.01833*}
} & \parbox{109pt}{\raggedright 
{\small Kariwa et al. \citep{46Kariwa}}
} \\
\hline
\parbox{88pt}{\raggedright 
{\small SARS-CoV-1}
} & \parbox{60pt}{\raggedright 
{\small 121}
} & \parbox{68pt}{\raggedright 
{\small log 5.325}
} & \parbox{68pt}{\raggedright 
{\small 0.01017}
} & \parbox{109pt}{\raggedright 
{\small Kariwa et al. \citep{46Kariwa}}
} \\
\hline
\parbox{88pt}{\raggedright 
{\small SARS-CoV-1}
} & \parbox{60pt}{\raggedright 
{\small 1000}
} & \parbox{68pt}{\raggedright 
{\small $\geq{}$log 4.81}
} & \parbox{68pt}{\raggedright 
{\small 0.00111}
} & \parbox{109pt}{\raggedright 
{\small Heimbuch \& Harnish \citep{34Heimbuch}}
} \\
\hline
\parbox{88pt}{\raggedright 
{\small SARS-CoV-1}
} & \parbox{60pt}{\raggedright 
{\small 50}
} & \parbox{68pt}{\raggedright 
{\small log 3.05}
} & \parbox{68pt}{\raggedright 
{\small 0.01405}
} & \parbox{109pt}{\raggedright 
{\small Eickmann et al. \citep{33Eickmann}}
} \\
\hline
\parbox{88pt}{\raggedright 
{\small SARS-CoV-1}
} & \parbox{60pt}{\raggedright 
{\small 100}
} & \parbox{68pt}{\raggedright 
{\small $\geq{}$log 3.5}
} & \parbox{68pt}{\raggedright 
{\small 0.00806}
} & \parbox{109pt}{\raggedright 
{\small Eickmann te al. \citep{33Eickmann}}
} \\
\hline
\parbox{88pt}{\raggedright 
{\small MERS-CoV}
} & \parbox{60pt}{\raggedright 
{\small 1000}
} & \parbox{68pt}{\raggedright 
{\small $\geq{}$log 4.5}
} & \parbox{68pt}{\raggedright 
{\small 0.00104}
} & \parbox{109pt}{\raggedright 
{\small Heimbuch \& Harnish \citep{34Heimbuch}}
} \\
\hline
\parbox{88pt}{\raggedright 
{\small SARS-CoV-2}
} & \parbox{60pt}{\raggedright 
{\small 3.7}
} & \parbox{68pt}{\raggedright 
{\small log 3.3}
} & \parbox{68pt}{\raggedright 
{\small 0.20536}
} & \parbox{109pt}{\raggedright 
{\small Bianco at el. \citep{42Bianco}}
} \\
\hline
\parbox{88pt}{\raggedright 
{\small SARS-CoV-2}
} & \parbox{60pt}{\raggedright 
{\small 5}
} & \parbox{68pt}{\raggedright 
{\small log 2.0}
} & \parbox{68pt}{\raggedright 
{\small 0.09210}
} & \parbox{109pt}{\raggedright 
{\small Signify \citep{55Signify}}
} \\
\hline
\parbox{88pt}{\raggedright 
{\small SARS-CoV-2}
} & \parbox{60pt}{\raggedright 
{\small 22}
} & \parbox{68pt}{\raggedright 
{\small log 6.0}
} & \parbox{68pt}{\raggedright 
{\small 0.06280}
} & \parbox{109pt}{\raggedright 
{\small Signify \citep{55Signify}}
} \\
\hline
\parbox{88pt}{\raggedright 
{\small SARS-CoV-2}
} & \parbox{60pt}{\raggedright 
{\small 3.75**}
} & \parbox{68pt}{\raggedright 
{\small log 0.9}
} & \parbox{68pt}{\raggedright 
{\small 0.05526}
} & \parbox{109pt}{\raggedright 
{\small Inagaki et al. \citep{40Inagaki}}
} \\
\hline
\parbox{88pt}{\raggedright 
{\small SARS-CoV-2}
} & \parbox{60pt}{\raggedright 
{\small 37.5**}
} & \parbox{68pt}{\raggedright 
{\small log 3.0}
} & \parbox{68pt}{\raggedright 
{\small 0.01842}
} & \parbox{109pt}{\raggedright 
{\small Inagaki et al. \citep{40Inagaki}}
} \\
\hline
\end{tabular}
\caption{Calculated \textit{Z} values for the UV-C (254\,nm) and deep-UV (280\,nm) irradiation experiments involving coronaviruses suspended in liquids \citep{Beggs2020}. \\ {\small * Estimated from plots and data presented in source material.} \\ {\small ** Using deep-UV light at 280\,nm (all other experiments performed using UV-C light at 254\,nm).} } \label{tab:CB1}
\end{table}
\vspace{2pt}
\end{center}

While no irradiation experiments have yet been undertaken on SARS-CoV-2 virions in air, a few aerosol experiments have been done using other related coronaviruses. The results of these experiments, together with other selected viruses for comparison purposes, are listed in table \ref{tab:CB2}. These reveal that compared to influenza A (mean \textit{Z} = 0.19435 m$^{2}$/J), the coronaviruses (\textit{Z} = 0.377 m$^{2}$/J for murine (mouse) hepatitis virus (MHV) coronavirus \citep{54Walker}; 0.410 m$^{2}$/J for human coronavirus 229E \citep{52Buonanno}; and 0.590 m$^{2}$/J for human coronavirus OC43 \citep{52Buonanno}) are much more susceptible to UV damage, which is perhaps only to be expected given that they have a genome that is approximately twice the length of the influenza virus \citep{51Sagripanti}. Importantly, the coronavirus \textit{Z} values are an order of magnitude greater than those obtained for SARS-CoV-2 in liquid, implying that when aerosolised, coronaviruses in general and SARS-CoV-2 in particular, are much easier to disinfect compared with when they are presented in liquids or on surfaces. This however is to be expected as the medium in which microbes are irradiated greatly influences the magnitude of the \textit{Z} value achieved. This is because UV-C light is attenuated, due to scattering and absorption \citep{56Gregory}, as it passes through liquids \citep{57Mamane}. Consequently, viruses are generally easier to disinfect in the air compare with when they are on surfaces or in liquids.\\
\\ 

\begin{table}
\begin{tabular}{|p{92pt}|p{116pt}|p{63pt}|p{63pt}|p{90pt}|}
\hline
\parbox{92pt}{\raggedright 
{\small Researchers}
} & \parbox{116pt}{\raggedright 
{\small Virus}
} & \parbox{63pt}{\raggedright 
{\small UV-C wavelength}
} & \parbox{63pt}{\raggedright 
{\small Effective Z value}

{\small (m$^{2}$/J)}
} & \parbox{90pt}{\raggedright 
{\small Reporter}
} \\
\hline
\parbox{92pt}{\raggedright 
{\small Jensen \citep{58Jensen}}
} & \parbox{116pt}{\raggedright 
{\small Adenovirus}
} & \parbox{63pt}{\raggedright 
{\small 254\,nm}
} & \parbox{63pt}{\raggedright 
{\small 0.0546}
} & \parbox{90pt}{\raggedright 
{\small Kowalski et al. \citep{59Kowalski}}
} \\
\hline
\parbox{92pt}{\raggedright 
{\small Jensen \citep{58Jensen}}
} & \parbox{116pt}{\raggedright 
{\small Coxsackie B-1}
} & \parbox{63pt}{\raggedright 
{\small 254\,nm}
} & \parbox{63pt}{\raggedright 
{\small 0.1108}
} & \parbox{90pt}{\raggedright 
{\small Kowalski et al. \citep{59Kowalski}}
} \\
\hline
\parbox{92pt}{\raggedright 
{\small Jensen \citep{58Jensen}}
} & \parbox{116pt}{\raggedright 
{\small Influenza A}
} & \parbox{63pt}{\raggedright 
{\small 254\,nm}
} & \parbox{63pt}{\raggedright 
{\small 0.1187}
} & \parbox{90pt}{\raggedright 
{\small Kowalski et al. \citep{59Kowalski}}
} \\
\hline
\parbox{92pt}{\raggedright 
{\small Jensen \citep{58Jensen}}
} & \parbox{116pt}{\raggedright 
{\small Sindbis virus}
} & \parbox{63pt}{\raggedright 
{\small 254\,nm}
} & \parbox{63pt}{\raggedright 
{\small 0.1040}
} & \parbox{90pt}{\raggedright 
{\small Kowalski \citep{60Kowalski}}
} \\
\hline
\parbox{92pt}{\raggedright 
{\small \citep{58Jensen}}
} & \parbox{116pt}{\raggedright 
{\small Vaccinia virus}
} & \parbox{63pt}{\raggedright 
{\small 254\,nm}
} & \parbox{63pt}{\raggedright 
{\small 0.1528}
} & \parbox{90pt}{\raggedright 
{\small Kowalski et al. \citep{59Kowalski}}
} \\
\hline
\parbox{92pt}{\raggedright 
{\small Walker \& Ko \citep{54Walker}}
} & \parbox{116pt}{\raggedright 
{\small Adenovirus}
} & \parbox{63pt}{\raggedright 
{\small 254\,nm}
} & \parbox{63pt}{\raggedright 
{\small 0.0390}
} & \parbox{90pt}{\raggedright 
{\small Walker \& Ko \citep{54Walker}}
} \\
\hline
\parbox{92pt}{\raggedright 
{\small Walker \& Ko \citep{54Walker}}
} & \parbox{116pt}{\raggedright 
{\small MHV coronavirus}
} & \parbox{63pt}{\raggedright 
{\small 254\,nm}
} & \parbox{63pt}{\raggedright 
{\small 0.3770}
} & \parbox{90pt}{\raggedright 
{\small Wolker \& Ko \citep{54Walker}}
} \\
\hline
\parbox{92pt}{\raggedright 
{\small McDevitt et al. \citep{53McDevitt}}
} & \parbox{116pt}{\raggedright 
{\small Influenza A}
} & \parbox{63pt}{\raggedright 
{\small 254\,nm}
} & \parbox{63pt}{\raggedright 
{\small 0.2700}
} & \parbox{90pt}{\raggedright 
{\small McDevitt et al. \citep{53McDevitt}}
} \\
\hline
\parbox{92pt}{\raggedright 
{\small McDevitt et al. \citep{61McDevitt}}
} & \parbox{116pt}{\raggedright 
{\small Vaccinia virus}
} & \parbox{63pt}{\raggedright 
{\small 254\,nm}
} & \parbox{63pt}{\raggedright 
{\small 2.5400}
} & \parbox{90pt}{\raggedright 
{\small McDevitt ct al. \citep{61McDevitt}}
} \\
\hline
\parbox{92pt}{\raggedright 
{\small Buonanno et al. \citep{52Buonanno}}
} & \parbox{116pt}{\raggedright 
{\small Human coronavirus 229E}
} & \parbox{63pt}{\raggedright 
{\small 222\,nm}
} & \parbox{63pt}{\raggedright 
{\small 0.4100}
} & \parbox{90pt}{\raggedright 
{\small Buonanno et al. \citep{52Buonanno}}
} \\
\hline
\parbox{92pt}{\raggedright 
{\small Buonanno et al. \citep{52Buonanno}}
} & \parbox{116pt}{\raggedright 
{\small Human coronavirus OC43}
} & \parbox{63pt}{\raggedright 
{\small 222\,nm}
} & \parbox{63pt}{\raggedright 
{\small 0.5900}
} & \parbox{90pt}{\raggedright 
{\small Buonanno et al. \citep{52Buonanno}}
} \\
\hline
\end{tabular}
\caption{Summary of reported effective \textit{Z} values for single-pass UV-C irradiation experiments performed on selected aerosolised viruses in air \citep{Beggs2020}} \label{tab:CB2}
\end{table}

Collectively, the results presented above strongly suggest that the SARS-CoV-2 virus is relatively easily inactivated by UV-C light and that when aerosolised the virus is likely to exhibit a UV susceptibility constant, \textit{Z}, that is similar in magnitude to other coronaviruses in air. As such, this indicates that SARS-CoV-2, when suspended in air, should be reasonably easy to inactivate using UV light at 254\,nm. As such, UVGI air disinfection applications appear to have potential as an intervention to inhibit the transmission of COVID-19 in buildings and other enclosed spaces. 

\subsection{UVGI air disinfection applications}

While it is clear from the discussion above that the SARS-CoV-2 virus can be relatively easily inactivated using UV-C light, this does not necessarily mean that installing UVGI air disinfection in buildings or other enclosed spaces (e.g. passenger vehicle, trains, buses, etc.) will prevent the transmission of COVID-19 in these facilities. This is because, in order to break the chain of COVID-19 transmission, UVGI needs to be applied in a manner that: (i) is appropriate to the situation; (ii) targets the route of transmission; and (iii) disinfects enough air to be effective. Too often do building owners and occupiers, seduced by adverts claiming a ‘99.9\% kill’ against an infectious pathogen, install expensive UVGI room air cleaners in their facilities, only to find that they offer little or no protection at all. This occurs because the impressive claims made by manufacturers often relate to single-pass microbiological tests (which may or may not involve relevant microbial species) rather than addressing how the device will actually perform in a given room space. So for example, a UVGI room air cleaner may inactivate (kill) every microbe that passes through it, but if the air flow rate through the device is small, then it will have little impact on a large, well ventilated, room space. Therefore, when considering UVGI air disinfection it is important to evaluate its likely performance in the context of the room space, something which will inevitably involve evaluation of room occupancy levels and the geometry of the room space, as well as consideration of the ventilation system.
   
UVGI air disinfection devices can be broadly classified into:
\begin{itemize}
\item \textbf{Upper-room UVGI systems} in which an open UV-C irradiation field above the heads of room occupants is used to disinfect aerosolised bacteria and viruses suspended in the air. Because UV-C light is harmful to humans, such systems utilize louvres and baffles that obscure the UV lamps from eyesight so that the room occupants are kept safe.
\item \textbf{In-duct UVGI systems} which utilise UV-C lamps mounted in the return or supply air ducts of mechanical HVAC systems to disinfect the air either to or from the room space.
\item \textbf{UVGI room air cleaners} which are located within the room space and employ UV-C lamps mounted in a container with a fan. These can vary in size, but only disinfect the air that passes through the device. 
\end{itemize}

A detailed discussion of each of the above UVGI variants follows in the sections below, so here we will restrict ourselves to a broad discussion of some general principles that are applicable to all types of UVGI air disinfection system.

\subsubsection{Target microbes}

Unlike air filters, which capture all particulates of a given size irrespective of their biological status, UVGI air disinfection systems target specific microbes. They do not clean the air; rather they use biophysical mechanisms to inactivate target viral and bacterial species, which once inactivated remain in the air stream. Target microbes are generally viral or bacterial species that cause infectious disease. So in the case of COVID-19, the target microbe is the SARS-CoV-2 virus. So when sizing a UVGI air disinfection system to inhibit the transmission of COVID-19, it is necessary to ensure that any SARS-CoV-2 virions in the air will receive a lethal dose of UV-C light, since a sub-lethal dose may leave some virions infectious. In order to calculate the necessary lethal dose to be administered, it is first necessary to use the appropriate \textit{Z} value for the target microbe in an aerosolised state, and then, if possible, add a factor of safety in order to ensure that the installation will adequately protect the room occupants. In the case of COVID-19, although the precise \textit{Z} value for SARS-CoV-2 in air is not known, \textit{Z} values for three other closely related coronaviruses in air have been determined \citep{52Buonanno,54Walker}. Of these, the \textit{Z} value (0.377 m$^{2}$/J) for MHV coronavirus \citep{54Walker} is the lowest value, implying that this is the most hardy of the three. It is therefore suggested that this is probably a good candidate for the target \textit{Z} value of SARS-CoV-2 in air \citep{Beggs2020} as it is likely to be a conservative value.

\subsubsection{Safety}

UV-C light is highly biologically active and as such can cause damage to humans. Furthermore, although UV light is far more energetic than the visible portion of the electromagnetic spectrum, it is invisible to humans, and so damage can occur without the individuals concerned noticing. Exposure to UV light can result in transient corneal inflammation (\textit{photokeratitis}) in the eye \citep{Cullan2002}, which may go unnoticed, but may progress to include inflammation of the conjunctiva (\textit{photoconjunctivitis}) \citep{Grifoni2005}. Acute exposure to UV-C radiation can cause more severe corneal damage \citep{Cullan2002}, which generally abates after several days, leaving no permanent damage \citep{Cullan2002}, although there have been reports of clinical symptoms persisting for as much a two years after an acute UV injury \citep{Zaffina2012}. For this reason, it is important to ensure that UV-C lamps are completely hidden from the view of room occupants, either through total enclosure or the use of louvres or shields. In addition, in order to protect the eyes, UV-C safety goggles should be worn when undertaking maintenance work on UV systems.

UV-C light can also cause cutaneous damage (erythema) \citep{Harrison2002}, resulting in reddening of the skin akin to sunburn. In this respect, UV-C light is actually less effective than UV-B at causing erythema because it has a lower penetration capability. More importantly, because prolonged exposure to UV light is associated with cancer, doubts have been raised about the safety of upper-room UVGI air disinfection systems due to the risk of exposure to reflected UV-C light. However, the International Commission on Illumination (CIE) review of the subject found that upper-room UVGI air disinfection could be safely used without significant risk for long-term delayed effects such as skin cancer \citep{63CIE}.

\subsubsection{What UVGI can and cannot do}
Irrespective of the specific type of UVGI air disinfection used, it is important to be aware of the limitations of the technology. While UVGI air disinfection can help to reduce the airborne (aerosol) transmission of infectious diseases, it cannot protect against close range droplet transmission. So in the context of COVID-19, this means that UVGI air disinfection cannot protect an individual if, for example, an infected person coughs directly in their face. However, if upper-room UVGI is applied, it can protect individuals from the many hundreds of smaller aerosol particles (generally $<$50 $\mu{}$m in diameter) that are exhaled by room occupants, which can be readily transported on room air currents \citep{18Beggs}. In addition, when mounted in return air ductwork, UVGI lamps can also protect against infectious aerosol particles being recirculated back into the room spaces through the mechanical ventilation system, if the system recirculates a portion of the air in order to save energy.

\subsection{Upper-room UVGI air disinfection}

\begin{figure}
\begin{center}
\includegraphics[width=249pt]{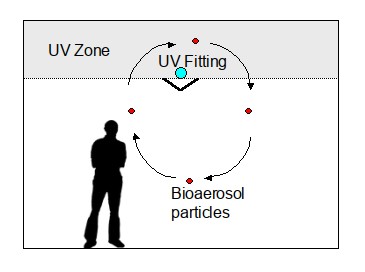}
\end{center}
\caption{An upper-room UVGI air disinfection installation} \label{fig:CB1}
\end{figure}

\begin{figure}
\begin{center}
\includegraphics[width=177pt]{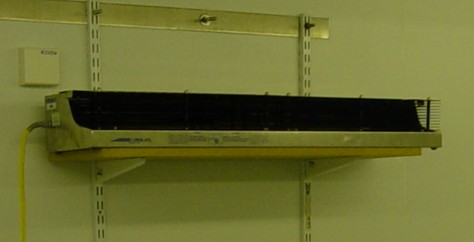}  \\

\end{center}
    \caption{Typical wall mounted upper-room UVGI air disinfection fitting.}
    \label{fig:CB2}
\end{figure}

Upper-room UVGI, as the name suggests, involves the creation of an open field UV-C irradiation field within the upper portion of a room as shown in figure \ref{fig:CB1}. This can be done using either wall or pendant fittings mounted at high level similar to those shown in figure \ref{fig:CB2}. Since exposure to UV-C light can be harmful, such devices are fitted with louvres or baffles, in order to prevent: (i) the occupants below from seeing the lamps; and (ii) the UV light from penetrating into the lower room space. The goal of upper room UVGI is to inactivate any infectious microbes (e.g. viruses and bacteria) that may pass through the UV zone and thus disinfect the air inhaled in the lower room space. As such, upper-room installations rely on natural convection currents, rather than fans, to carry aerosol particles through the UV field. By using the convection currents that occur naturally in room spaces it is possible to disinfect very large volumes of air relatively quickly \citep{64Miller}. As such, upper-room UVGI air disinfection is a well-established technology \citep{65First,66First} that has proven effective as a public health intervention to prevent the spread of airborne diseases such as measles \citep{67Nardell} and tuberculosis (TB) in buildings \citep{Escombe2009,69Noakes}. 

\subsubsection{Calculating the required UV flux}

Despite first being used in the 1930s \citep{39Reed}, the guidelines for designing upper-room UVGI systems are surprisingly sparse. This is primarily because the required UV-C radiant flux, \textit{E}, is wholly dependent on the target microbe and the air movement characteristics of the room space, both of which may vary greatly depending on the application being considered. Also, because most UVGI air disinfection activity has focused on the eradication of TB \citep{coker2001guidelines}, the guidelines which exist tend to be written in the context of that disease and generally take a ‘rule-of-thumb’ approach. With respect to this, Table \ref{tab:CB3}, which is taken from the ASHRAE UVGI air disinfection guidelines \citep{62ASHRAE}, is typical of the genre.

\begin{center}

\vspace{3pt} \noindent

\begin{table}
\begin{tabular}{|p{95pt}|p{70pt}|p{71pt}|p{71pt}|p{71pt}|}
\hline
\parbox{95pt}{\raggedright } & \parbox{70pt}{\raggedright 
{\scriptsize Wall-Mounted Fixtures}
} & \parbox{71pt}{\raggedright 
{\scriptsize Wall-Mounted Fixtures}
} & \parbox{71pt}{\raggedright 
{\scriptsize Ceiling-Mounted Fixtures}
} & \parbox{71pt}{\raggedright 
{\scriptsize Ceiling-Mounted Fixtures}
} \\
\hline
\parbox{95pt}{\raggedright } & \parbox{70pt}{\raggedright 
{\scriptsize Corner Mount}
} & \parbox{71pt}{\raggedright 
{\scriptsize Wall Mount}
} & \parbox{71pt}{\raggedright 
{\scriptsize Pendent}
} & \parbox{71pt}{\raggedright 
{\scriptsize Pendent with Fan}
} \\
\hline
\parbox{95pt}{\raggedright 
{\scriptsize Beam pattern}
} & \parbox{70pt}{\raggedright 
{\scriptsize 90$^\circ{}$}
} & \parbox{71pt}{\raggedright 
{\scriptsize 180$^\circ{}$}
} & \parbox{71pt}{\raggedright 
{\scriptsize 360$^\circ{}$}
} & \parbox{71pt}{\raggedright 
{\scriptsize 360$^\circ{}$}
} \\
\hline
\parbox{95pt}{\raggedright 
{\scriptsize Minimum ceiling height}
} & \parbox{70pt}{\raggedright 
{\scriptsize 2.44 m}
} & \parbox{71pt}{\raggedright 
{\scriptsize 2.44 m}
} & \parbox{71pt}{\raggedright 
{\scriptsize 2.89 m}
} & \parbox{71pt}{\raggedright 
{\scriptsize 2.89 m}
} \\
\hline
\parbox{95pt}{\raggedright 
{\scriptsize Fixture mounted height}
} & \parbox{70pt}{\raggedright 
{\scriptsize 2.1 m}
} & \parbox{71pt}{\raggedright 
{\scriptsize 2.1 m}
} & \parbox{71pt}{\raggedright 
{\scriptsize 2.4 m}
} & \parbox{71pt}{\raggedright 
{\scriptsize 2.4 m}
} \\
\hline
\parbox{95pt}{\raggedright 
{\scriptsize Ideal UV-C intensity (flux) for effective disinfection}
} & \parbox{70pt}{\raggedright 
{\scriptsize $>$ 10 $\mu{}$W/cm$^{2}$}
} & \parbox{71pt}{\raggedright 
{\scriptsize $>$ 10 $\mu{}$W/cm$^{2}$}
} & \parbox{71pt}{\raggedright 
{\scriptsize $>$ 10 $\mu{}$W/cm$^{2}$}
} & \parbox{71pt}{\raggedright 
{\scriptsize $>$ 10 $\mu{}$W/cm$^{2}$}
} \\
\hline
\end{tabular}
\vspace{2pt}
\caption{Suggested installation summary for upper-room UVGI air disinfection \citep{62ASHRAE,coker2001guidelines}} \label{tab:CB3}
\end{table}
\end{center}

While to date no guidelines exist regarding upper-room UVGI and COVID-19, Beggs and Avital recently produced a feasibility study \citep{Beggs2020} to evaluate the potential efficacy of the technology in this context. This study included a methodology (presented below) for estimating the upper-room UV flux required to disinfect the SARS-CoV-2 virus in ventilated room spaces. This approach utilized the methodology described in Beggs and Sleigh \citep{71Beggs} and assumed that the room air is well mixed, which is a reasonable approximation for many applications \citep{71Beggs}. If the space is well mixed, then the average particle residence time, \textit{t$_{res}$}, (in seconds) in the room will be
\begin{equation}
t_{res}=\frac{1}{n}\times{3600}  \, , \label{eq:CB5}    
\end{equation}
where \textit{n} is the room ventilation rate in air changes per hour (AC/hr).

From equation \ref{eq:CB5} it can be approximated that the average particle residence time in the upper-room UV field, \textit{t$_{uv}$}, (in seconds) will be
\begin{equation}
t_{uv}=t_{res}\times{\frac{h_{uv}}{h_r}}  \, , \label{eq:CB6}    
\end{equation}
where \textit{h$_{r}$} is the floor-to-ceiling height (m), and \textit{h$_{uv}$} is the depth of the upper-room UV zone (m).

Because \textit{Z} values are often determined experimentally using microbes suspended in liquids or on surfaces, it may by necessary to adjust the \textit{Z} value for use with upper-room UVGI systems \citep{72Beggs,73Yang}, as follows
\begin{equation}
Z_{ur}=Z\times{c_{ur}}  \, , \label{eq:CB7}    
\end{equation}
where \textit{Z$_{ur}$} is the effective upper-room \textit{Z} value (m$^{2}$/J), and \textit{c$_{ur}$} is a correction coefficient.

\textit{Z$_{ur}$} can be assumed to be the same as the aerosol \textit{Z} value for the target microbe, examples of which are presented in table \ref{tab:CB2}.

So if we assume that the air in a room is well mixed, by combining equations \ref{eq:CB2}, \ref{eq:CB3} and \ref{eq:CB6} it is possible to compute the average irradiation flux (W/m$^{2}$),
\textit{E$_{r}$}, that is required to achieve a desired survival fraction, \textit{f}$_{r}$.
\begin{equation}
E_r=\frac{1}{(Z_{ur}\times{t_{uv}})}\times{ln(f_r)}  \, . \label{eq:CB8}    
\end{equation}
In which case, the average UV dose received (J/m$^{2}$), \textit{H}$_{r}$, is
\begin{equation}
H_r=E_r\times{t_{uv}}  \, . \label{eq:CB9}    
\end{equation}
Alternatively, the disinfection achieved by an upper-room UVGI system can be thought of as being equivalent to additional air changes in the room space \citep{74McDevitt}. In this scenario, the UV rate constant, \textit{k$_{uv}$}, which can be thought of as the equivalent air change rate per second, can be determined using \citep{72Beggs}, i.e. 
\begin{equation}
k_{uv}=Z_{ur}\times{E}\times{\frac{h_{uv}}{h_r}}  \, . \label{eq:CB10}    
\end{equation}

So in a ventilated room in which contamination ceases at time zero, we can utilize both the UV rate constant, \textit{k$_{uv}$}, and a rate constant,
\textit{k$_{v}$}, for the ventilation (i.e. \textit{n} $\div{}$ 3600), to produce
a decay model for the room space
\begin{equation}
C_t=C_0\times{e^{(k_{v}+k_{uv}+k_{d}})}  \, ,\label{eq:CB11}
\end{equation}
where \textit{C$_{0}$} and \textit{C$_{t}$} are the concentrations of viable
viral particles in the room space (virions/m$^{3}$) at time zero and \textit{t}
seconds respectively; \textit{k$_{v}$} is the ventilation rate constant;
\textit{k$_{d}$} is the particulate deposition rate constant (e.g. 0.0014
s$^{-1}$ \citep{stadnytskyi2020airborne}); and \textit{t }is time in seconds.

Similarly, the following continuous contamination model represents the
contaminant concentration in the room space, \textit{C$_{uv}$}, under
steady-state conditions
\begin{equation}
C_{uv}=\frac{q}{(k_v+k_{uv}+k_d)\times{V})}  \, , \label{eq:CB12}
\end{equation}
where \textit{V} is the room volume (m$^{3}$), and \textit{q} is the steady-state room contamination rate (virions/s).

Using this simple approach, it is possible to estimate:
\begin{itemize}
	\item The average UV-C flux ($\mu{}$W/cm$^{2}$), \textit{E$_{r}$}, required to achieve
the desired level of inactivation in the target microbe.
	\item The average UV-C dose (mJ/cm$^{2}$), \textit{H$_{r}$}, required to achieve the
desired level of inactivation in the target microbe.
	\item The equivalent air change rate per second, \textit{k$_{uv}$}, that can be
achieved by the UVGI installation with regard to the target microbe.
\end{itemize}

\subsubsection{Upper-room UVGI feasibility study}

Recently Beggs and Avital \citep{Beggs2020} undertook a feasibility study to evaluate the potential efficacy of the upper-room UVGI air disinfection as a measure to prevent the transmission of COVID-19 in a ventilated room space. In this study equations \ref{eq:CB6} and \ref{eq:CB8} were used to estimate the average upper-room irradiation flux that would be required to achieve a 50 - 90\% reduction in aerosolised SARS-CoV-2 virions (through the action of the UV-C alone) in a 4.2 $\times{}$ 4.2 $\times{}$ 2.5 m high room space for a range of ventilation rates. These dimensions were chosen because they are typical for an upper-room UVGI installation in which the lamp height is 2.1 m above the floor \citep{66First}. In the model it was assumed that the air was completely mixed, which meant that according to equation \ref{eq:CB6}, aerosol particles would spend on average 16\% of their room residency time in the UV zone.

In addition to computing the required UV flux, the performance of a standard upper-room UVGI fitting was evaluated against the challenge of SARS-CoV-2. This was done in accordance with the guidelines stated by First \citep{66First}, in which it was assumed that the room contained a single 30 W (input) UV-C fitting capable of delivering an average upper-room flux of 50 $\mu{}$W/cm$^{2}$ \citep{62ASHRAE}, with performance modelled in terms of equivalent ventilation rate using equation \ref{eq:CB10}. 

Because to date no UV irradiation experiments have been performed on aerosols containing the SARS-CoV-2 virus, in the feasibility study the value of \textit{Z$_{ur}$} was taken to be 0.377 m$^{2}$/J, which was the value that Walker and Ko obtained for the MHV coronavirus in air \citep{54Walker}. Because this was considered a conservative value, it was selected as a suitable surrogate for SARS-CoV-2. In addition, because of the uncertainty associated with this assumed value, a 10-fold ‘factor of safety’ was introduced into the analysis by also modelling a worst-case scenario in which \textit{Z$_{ur}$} was 0.0377 m$^{2}$/J. 

Table \ref{tab:CB4} presents the results of the room analysis using these two values for \textit{Z$_{ur}$}, for a range of ventilation rates. From this it can be seen that there is a direct inverse relationship between particle residence time in the UV field,  \textit{t$_{uv}$}, and the required irradiation flux, \textit{E$_{r}$}, as predicted by equation \ref{eq:CB8}. This means that for any given \textit{Z} value, the value of \textit{E$_{r}$} will double as the room ventilation rate doubles. The table also reveals that there is a direct inverse relationship between \textit{Z$_{ur}$} and \textit{E$_{r}$}. From the calculated values in this table it can be seen that if \textit{Z$_{ur}$} = 0.377 m$^{2}$/J, then with an average UV flux of just 10 $\mu{}$W/cm$^{2}$ it should be possible to achieve $>$90\% inactivation of SARS-CoV-2, even at a ventilation rate of 8 AC/h. However, if in reality, \textit{Z$_{ur}$}, is 0.0377 m$^{2}$/J, then all the calculated fluxes would have to increase by a factor of ten to achieve the same results. Given that accepted guidelines \citep{66First} recommend for a room 2.5 m high, one 30 W (input) UV lamp per 18.58 m$^{2}$ of floor area, which will produce an average flux in the region 50 $\mu{}$W/cm$^{2}$, this means that even under this worst-case scenario it should still be possible to achieve disinfection rates $>$90\% for all but the highest ventilation rates.

When the UV flux was fixed at an average of 50 $\mu{}$W/cm$^{2}$, it was found that for \textit{Z$_{ur}$}, = 0.377 m$^{2}$/J the upper-room UVGI installation produced an equivalent air change rate of 108.6 AC/h, whereas if \textit{Z$_{ur}$}, = 0.0377 m$^{2}$/J this fell to 10.9 AC/h. These values were constant and unaffected by the actual room ventilation rate. 

\begin{table}
\vspace{3pt} \noindent
\begin{tabular}{|p{76pt}|p{58pt}|p{58pt}|p{66pt}|p{66pt}|p{67pt}|}
\hline
\parbox{76pt}{\raggedright 
{\small Ventilation rate}

{\small (AC/h)}
} & \parbox{58pt}{\raggedright 
{\small Average particle residence time in UV field.}

{\small (mins.)}
} & \parbox{58pt}{\raggedright 
{\small UV susceptibility constant, \textit{Z$_{ur}$}}

{\small (m$^{2}$/J)}
} & \parbox{66pt}{\raggedright 
{\small Average irradiance required for 50\% inactivation}

{\small ($\mu{}$W/cm$^{2}$)}
} & \parbox{66pt}{\raggedright 
{\small Average irradiance required for 70\% inactivation}

{\small ($\mu{}$W/cm$^{2}$)}
} & \parbox{67pt}{\raggedright 
{\small Average irradiance required for 90\% inactivation}

{\small ($\mu{}$W/cm$^{2}$)}
} \\
\hline
\parbox{76pt}{\raggedright 
{\small 1}
} & \parbox{58pt}{\raggedright 
{\small 9.6}
} & \parbox{58pt}{\raggedright 
{\small 0.3770}
} & \parbox{66pt}{\raggedright 
{\small 0.319}
} & \parbox{66pt}{\raggedright 
{\small 0.554}
} & \parbox{67pt}{\raggedright 
{\small 1.060}
} \\
\hline
\parbox{76pt}{\raggedright 
{\small 2}
} & \parbox{58pt}{\raggedright 
{\small 4.8}
} & \parbox{58pt}{\raggedright 
{\small 0.3770}
} & \parbox{66pt}{\raggedright 
{\small 0.638}
} & \parbox{66pt}{\raggedright 
{\small 1.109}
} & \parbox{67pt}{\raggedright 
{\small 2.121}
} \\
\hline
\parbox{76pt}{\raggedright 
{\small 4}
} & \parbox{58pt}{\raggedright 
{\small 2.4}
} & \parbox{58pt}{\raggedright 
{\small 0.3770}
} & \parbox{66pt}{\raggedright 
{\small 1.277}
} & \parbox{66pt}{\raggedright 
{\small 2.218}
} & \parbox{67pt}{\raggedright 
{\small 4.241}
} \\
\hline
\parbox{76pt}{\raggedright 
{\small 6}
} & \parbox{58pt}{\raggedright 
{\small 1.6}
} & \parbox{58pt}{\raggedright 
{\small 0.3770}
} & \parbox{66pt}{\raggedright 
{\small 1.915}
} & \parbox{66pt}{\raggedright 
{\small 3.327}
} & \parbox{67pt}{\raggedright 
{\small 6.362}
} \\
\hline
\parbox{76pt}{\raggedright 
{\small 8}
} & \parbox{58pt}{\raggedright 
{\small 1.2}
} & \parbox{58pt}{\raggedright 
{\small 0.3770}
} & \parbox{66pt}{\raggedright 
{\small 2.554}
} & \parbox{66pt}{\raggedright 
{\small 4.436}
} & \parbox{67pt}{\raggedright 
{\small 8.482}
} \\
\hline
\parbox{76pt}{\raggedright 
{\small 1}
} & \parbox{58pt}{\raggedright 
{\small 9.6}
} & \parbox{58pt}{\raggedright 
{\small 0.0377}
} & \parbox{66pt}{\raggedright 
{\small 3.192}
} & \parbox{66pt}{\raggedright 
{\small 5.544}
} & \parbox{67pt}{\raggedright 
{\small 10.604}
} \\
\hline
\parbox{76pt}{\raggedright 
{\small 2}
} & \parbox{58pt}{\raggedright 
{\small 4.8}
} & \parbox{58pt}{\raggedright 
{\small 0.0377}
} & \parbox{66pt}{\raggedright 
{\small 6.384}
} & \parbox{66pt}{\raggedright 
{\small 11.088}
} & \parbox{67pt}{\raggedright 
{\small 21.207}
} \\
\hline
\parbox{76pt}{\raggedright 
{\small 4}
} & \parbox{58pt}{\raggedright 
{\small 2.4}
} & \parbox{58pt}{\raggedright 
{\small 0.0377}
} & \parbox{66pt}{\raggedright 
{\small 12.768}
} & \parbox{66pt}{\raggedright 
{\small 22.177}
} & \parbox{67pt}{\raggedright 
{\small 42.414}
} \\
\hline
\parbox{76pt}{\raggedright 
{\small 6}
} & \parbox{58pt}{\raggedright 
{\small 1.6}
} & \parbox{58pt}{\raggedright 
{\small 0.0377}
} & \parbox{66pt}{\raggedright 
{\small 19.152}
} & \parbox{66pt}{\raggedright 
{\small 33.266}
} & \parbox{67pt}{\raggedright 
{\small 63.621}
} \\
\hline
\parbox{76pt}{\raggedright 
{\small 8}
} & \parbox{58pt}{\raggedright 
{\small 1.2}
} & \parbox{58pt}{\raggedright 
{\small 0.0377}
} & \parbox{66pt}{\raggedright 
{\small 25.536}
} & \parbox{66pt}{\raggedright 
{\small 44.355}
} & \parbox{67pt}{\raggedright 
{\small 84.829}
} \\
\hline
\end{tabular}
\vspace{2pt}
\caption{Predicted average upper-room UV irradiance fluxes required to achieve 50\%, 70\% and 90\% inactivation for SARS-CoV-2 assuming a range of \textit{Z$_{ur}$} values and ventilation rates. (Assuming \textit{Z$_{ur}$} = 0.377 or 0.0377 m$^{2}$/J) \citep{Beggs2020}} \label{tab:CB4}
\end{table}

Collectively, the results from the feasibility study suggest that upper-room UVGI may have considerable potential as an intervention against the transmission of COVID-19 in buildings, especially in situations where achieving high ventilation rates might otherwise be impractical. 

\subsubsection{Room air movement and particle decoupling}

While the results of the feasibility study are encouraging, it is important to remember that, unlike TB which is spread via the inhalation of droplet nuclei $<$5-10 $\mu{}$m in diameter, COVID-19 can be transmitted through the exhalation of larger respiratory droplets $<$100 $\mu{}$m, which rapidly reduce in size due to evaporation \citep{18Beggs,Xie2007,liu2017evaporation} to become aerosols say $<$50 $\mu{}$m in diameter \citep{21Nicas}. These aerosol particles have settling velocities $<$0.1 m/s and as such can readily be transported on convective room air currents, with the result that they can remain suspended in room air for many minutes. However, if the velocities of the convection currents drop, then some of the larger aerosol particles may decouple from the air stream and settle out due to gravitational deposition, potentially passing through the breathing zone where they can be inhaled by the room occupants. This is particularly the case if the air is poorly mixed, and stagnant regions exist within the room space. Under such circumstances larger aerosol particles may be inhaled without being fully irradiated by the upper-room UV field, undermining the effectiveness of the whole UVGI installation. Consequently, if upper-room UVGI is to be effective against COVID-19, it is important both to promote good room air mixing and also to ensure that larger aerosol particles (e.g. 10-50 $\mu{}$m in diameter) receive a lethal UV irradiation dose. As such, this may require upper-room UVGI systems to be supplemented with ceiling mounted fans \citep{76Zhu} or other devices to promote the necessary air movement to ensure that larger aerosol particles are adequately irradiated.

Upper-room UVGI air disinfection is highly dependent on good air mixing occurring between the upper and lower portions of the room space \citep{71Beggs,77Noakes,78Nicas}. In the feasibility study it was assumed that complete mixing occurred, which although a reasonable approximation in many instances, is not always the case because short circuiting can occur \citep{71Beggs}. If the room air mixing factor, which describes the inter-zonal air flow rate relative to the absolute room ventilation rate, is low, say for example due to stratification in a poorly ventilated space, then this can greatly impair the disinfection performance of an upper-room UVGI system \citep{71Beggs,77Noakes}. It is therefore important when designing such systems to carefully consider the air movement in the room space, in order to eliminate stagnant regions and maximise air movement through the UV field. With this in mind analysis can be performed using either zonal models \citep{77Noakes,79Noakes} or computational fluid dynamics (CFD) \citep{76Zhu,80Gilkeson}. Zonal models are relatively simple and easy to use, but limited in scope \citep{79Noakes}, whereas CFD is much more flexible and comprehensive, but is computationally expensive, requiring specialist skills and software to execute any models that are constructed.

\subsection{In-duct UVGI air disinfection} \label{sec:InDuct}

Many older commercial HVAC and mechanical ventilation systems recirculate a portion (in the region 50 -- 80\%) of the room air in order to save energy. In the context of the COVID-19 pandemic this strategy is potentially extremely hazardous because it means that aerosol particles containing the SARS-CoV-2 virus extracted from one location in a building may become widely distributed throughout the whole building by the mechanical (recirculating) ventilation system. If this happens, then the HVAC system could in effect become a pathogen distribution system. However, although in theory infectious aerosol particles can be dispersed around buildings by this route, the clinical evidence in support of this opinion is somewhat lacking, with for example, no evidence found to link the outbreak that occurred on the Diamond Princess cruise ship in January 2020 with the ship’s central air conditioning system \citep{81Xu}. Nevertheless, the UK \textit{Health and Safety Executive} (HSE) recommends:
\textit{“You can continue using most types of air conditioning system as normal. But, if you use a centralised ventilation system that removes and circulates air to different rooms it is recommended that you turn off recirculation and use a fresh air supply.”} \citep{82HSE}
and the Who Health Organization (WHO) states:
\textit{“For mechanical systems, increase the percentage of outdoor air, using economizer modes of HVAC operations and potentially as high as 100\%.”} \citep{83WHO}
While there may be controversy as to whether or not recirculating HVAC systems contribute to the spread of COVID-19, there is strong clinical evidence implicating recirculated air in the transmission of TB, with an outbreak on a US naval vessel attributed to a recirculating mechanical ventilation system, in which a single undiagnosed index case managed to infect 140 out of 308 crew members \citep{84Houk}. Given the evidence from this case and the general lack of knowledge surrounding COVID-19 transmission in buildings, it is therefore entirely understandable that the public health authorities might wish to adopt a precautionary approach in relation to the recirculation of return air.

Notwithstanding the above discussion, while it is possible in the UK to operate HVAC systems using 100\% outside air during the summer months, the situation is very different in the winter months when temperatures drop. This is because the heating coils in the air handling units (AHUs) in many installations do not have enough power to enable the supply air to be heated from, for example, --1$^{o}$C to say 28$^{o}$C. Consequently, if 100\% outside air is supplied, then during periods in which the weather is cold, many buildings might become uncomfortable and potentially uninhabitable. Consequently, making such buildings both COVID-19 ‘safe’ and also comfortable is a major challenge. However, it is this precise challenge to which in-duct UVGI air disinfection is particularly well suited. By installing UVGI lamps in the return air duct-work it is possible to irradiate the target microbes and inactivate them before they are recirculated back into the room space. In so doing, the UVGI lamps can in theory protect both the room occupants and the AHU from cross-infection/contamination, while still enabling the system to recirculate return air. So in the context of the COVID-19 pandemic, in-duct UVGI might be technology that could be retrofitted into commercial HVAC systems in order to enable buildings to become fully operational during the winter months.
 
In-duct UVGI air disinfection is also applicable to HVAC systems where thermal wheels are installed to recover waste heat. While these systems negate the need for the recirculation of return air, there is still the risk of cross-contamination with the SARS-CoV-2 virus because the wheel revolves between the two air streams. Employing UVGI lamps in the return air stream before the wheel potentially might prevent this from happening.

\subsubsection{Where should in-duct UVGI lamps be placed?}

In the context of the transmission of airborne disease (e.g. TB), while UVGI lamps can be placed in either the return or supply air ducts, it is generally preferable to install the lamps on the return air side of the AHU. This is because in this position, the UV field not only protects against pathogens being recirculated, but also prevents the up-stream duct-work, filters and coils from becoming contaminated. This is particularly important in the context of COVID-19 because coronaviruses can remain viable on surfaces for several days \citep{VanDoremalen2020} and therefore may pose a threat to staff performing maintenance duties. In addition, mounting the lamps on the return air duct before the AHU will also protect anyone in the vicinity of the system exhaust air outlet. 

\subsubsection{Failing safe}

With all UVGI systems it is important that they ‘fail safe’, so that building occupants are protected should the lamps fail. For this reason ASHRAE recommend that in-duct UVGI should \textit{“always be used in combination with proper filtration”} \citep{62ASHRAE}. In the context of COVID-19, this presents a tricky problem. Because COVID-19 is caused by a virus, this would necessitate the installation of HEPA filters, which would completely undermine the reason for retrofitting UVGI, namely to protect the building occupants while keeping fan energy consumption low. Therefore, workable solutions need to be developed to overcome this problem. 

\subsubsection{Computation of the UV flux}

While in theory the UV-C dose received by a microbe passing through a UV field can be easily computed using equation \ref{eq:CB2}, in practice with in-duct systems, computation of the dose received can be complex. Because the air velocities in mechanical ventilation ducts are relatively high, generally in the region 3-6 m/s, it means that residence times in the UV field are very short, generally $<$1 s. This means that in order to administer a lethal dose the UV flux has to be very large, often requiring multiple high-powered lamps. Consequently, without knowing the precise arrangement of the lamps and access to numerical modelling techniques it is difficult to compute the dose that will be received by a target microbe. The situation is further complicated by the fact that if the UV flux is not evenly distributed across the duct, microbes passing through the periphery of the field may not receive a lethal dose. If this occurs, the microbe may remain infectious, undermining the efficacy of the whole installation. As such, in order to properly evaluate how a given in-duct UVGI air disinfection installation will perform, it is necessary to employ CFD analysis \citep{85Capetillo,capetillo2017cfd,87Yang}. Although a rough estimate can be obtained of how a proposed in-duct UVGI installation might perform, using standard mathematical techniques \citep{88Beggs}, without CFD or some other advanced numerical modelling technique, it is not possible to be fully confident about any predictions made, unless of course experimental work is carried out. 

Given the complexity associated with designing in-duct UVGI air disinfection systems, it is perhaps not surprising that few guidelines exist regarding this type of installation \citep{62ASHRAE,89Kowalski}, with those that do exist being somewhat vague. For example, regarding in-duct UVGI the ASHRAE guidelines simply state:
\textit{“The required average irradiance for a typical in-duct system is on the order of 1000 to 10,000 $\mu{}$W/cm$^{2}$, but it could be higher or lower depending on the application requirements.”} \citep{62ASHRAE}
Because it is difficult for building owners and consultant engineers to perform the necessary in-duct UVGI calculations, they are forced to rely on the claims made by manufacturers regarding the performance of their equipment. Consequently, there is need for the manufacturers of UVGI air disinfection equipment to undertake robust microbiological testing, as well as fully characterising the UV fields produced by their devices.

\subsection{UVGI room air cleaners}

Many manufacturers make ‘UV-C in a box’ style air disinfection devices for use in room spaces. These devices essentially comprise UV-C lamps or LEDs mounted in a container with a fan. They are generally free-standing units designed to disinfect all the air that passes through the device. As such, manufacturers frequently make claims akin to ‘the unit achieves a 99.9\% disinfection rate’ based on a ‘single-pass’ microbiological test. These claims can be however extremely misleading, because they relate purely to the air that passes through the UV device and not to the effect that the unit will have on the room space, which can be calculated for steady-state using
\begin{equation}
C=\frac{q}{\dot{v}_r+(\dot{v}_{uv}\times{\eta})} \, , \label{eq:CB13}
\end{equation}
where \textit{C } is the contaminant concentration in the room space under steady-state conditions (e.g. (virions/m$^{3}$), \textit{q} is the steady-state room contamination rate (virions/s), $\dot{v}_r$ is the room ventilation rate (m$^{3}$/s), $\dot{v}_{uv}$  is the air flow rate through the UV air cleaner (m$^{3}$/s), and \textit{$\eta{}$} is the single-pass efficiency of the air disinfection device expressed as a fraction. 

From equation \ref{eq:CB13} it can be seen that if the room ventilation rate is large in comparison to the amount of air flowing through the UV air cleaner, then even if all the air passing through the unit is 100\% disinfected, the overall impact of the device on the room space will be minimal. So unless the air cleaner delivers a high air flow rate, or alternatively the room is very poorly ventilated, devices like this are unlikely to have a major impact on the transmission of COVID-19 when employed in large spaces.

Notwithstanding the discussion above, if used in small poorly ventilated spaces, or in conjunction with partitions, then small UVGI room air cleaners may be effective in protecting individuals against COVID-19. Consider for example, a communal office desk that has been partitioned with screens (i.e. creating booths) in order to minimise spread of COVID-19. If local air cleaning devices could be developed that disinfected the air within the individual spaces created by the desk partitions, then this might provide additional protection to individual workers. However, the feasibility of this would need to be verified either through CFD analysis or experimental work.

\subsection{Discussion}

The principal findings of this Appendix are that:
\begin{itemize}
	\item 	The SARS-CoV-2 virus is susceptible to damage from UV-C light and as such can be relatively easily inactivated using UVGI.
	\item	The SARS-CoV-2 virus appears to be particularly vulnerable to UV-C light when aerosolised and as such, may (in theory) be readily disinfected using upper-room UVGI air disinfection.
	\item	If retrofitted to recirculating HVAC installations, in-duct UVGI air disinfection may potentially be a useful technology for enabling commercial buildings to function during the winter months of the COVID-19 pandemic. 
	\item	With respect to COVID-19 transmission, no suitable guidelines exist regarding UVGI air disinfection.
\end{itemize}

When considering the above findings it is important to note the caveats \enquote{in theory} and \enquote{potentially}. These have been inserted because although there is strong evidence that the SARS-CoV-2 virus is susceptible to UV-C light, suggesting that UVGI air disinfection has considerable potential, there still is much doubt about how the technology should be applied in order to protect the occupants of buildings. While this doubt stems, in part, from an incomplete understanding of the mechanisms by which COVID-19 is transmitted, it is also due to inherent issues associated with the technology itself, which make its application difficult, as evidenced by the lack of detailed guidelines on the design of UVGI air disinfection systems. These inherent drawbacks relate specifically to: (i) the fact that the design of UVGI air disinfection systems is strongly influenced by the choice of the target microbe; and (ii) difficulty in predicting how any given installation will behave without the use of complex simulation tools like CFD. This makes it difficult to produce robust guidelines that are applicable to all situations. For example, the guidelines applicable to, say, the transmission of TB in hospitals, might not be suitable for protecting against the spread of COVID-19 in bars and music venues. Furthermore, once an air disinfection system is installed it is difficult, without extensive microbiological testing or a clinical trial, to tell whether or not the air disinfection system is actually protecting the occupants against infection. Collectively, these factors have greatly inhibited the uptake of UVGI air disinfection, with the result that many are suspicious of the technology.

Notwithstanding the above discussion, the results of the feasibility study by Beggs and Avital \citep{Beggs2020} suggest that upper-room UVGI air disinfection may well have a role to play in preventing the transmission of COVID-19 in some settings. However, much remains unknown about the limits of this technology with regard to COVID-19. Upper-room UVGI was primarily developed as an intervention against ‘true airborne’ diseases such as TB, where the infectious particles are $<$5-10 $\mu{}$m in diameter. However, COVID-19 is not a ‘true airborne’ disease in the classical sense, insomuch as it can also be transmitted by inhalation of aerosol droplets in the range 10-100 $\mu{}$m. Depending on their size and room air velocities, these larger aerosol particles can readily become suspended in the air and be dispersed around the room space. While many of these airborne particles are transported by convection currents to the upper part of the room space, some particles, due to their mass, may decouple from convective air streams and settle out due to gravitational deposition. In doing so they may fall through the breathing zone, where if they have not been fully inactivated by the UV field, they will be a potential hazard. So while it is undoubtedly the case that upper-room UVGI can help to inhibit the spread of some COVID-19, the extent to which it can be effective is not known and there is urgent need for CFD analysis work to evaluate how the technology can be adapted to be effective against COVID-19. In particular, there is a need to explore the use of room mounted fans in conjunction with UVGI air disinfection to promote better air movement between upper and lower room zones.

Given the urgent need for a solution, which will make commercial properties (offices, shops, pubs, etc.) and public buildings (schools, universities, etc.) COVID-19 safe as well as habitable during the winter months, perhaps the most promising potential application of UVGI air disinfection is as a retrofit for recirculating HVAC systems. By installing in-duct UVGI in the return air duct-work it should be possible to permit the recirculation of air while ensuring that SARS-CoV-2 virions are not recirculated. However, while this retrofit solution should not increase fan power consumption, the installation of UV-C lamps might considerably increase electrical power consumption, something which may make UVGI unfeasible. In addition, any requirement for ‘fail safe’ filters might also make in-duct UVGI untenable.

Because the performance of UVGI air disinfection installations is difficult to validate, architects, engineers and building owners are generally forced to rely on claims made by equipment manufacturers. However, the evidence base on which these claims are made, particularly those relating COVID-19, is often very weak, with little or no microbiological testing involving viruses undertaken. Consequently, commissioning bodies such as the NHS and government departments are reluctant to accept these claims without substantial clinical (biological) evidence, thus inhibiting the general uptake of UVGI air disinfection. As a result, there is urgent need for manufacturers of UVGI systems to undertake appropriate microbiological testing of their products so that they can demonstrate their efficacy against the SARS-CoV-2 virus.

Regarding UVGI air disinfection, the guidelines that exist \citep{62ASHRAE,65First,66First,coker2001guidelines}, which were largely developed to control the spread of TB \citep{coker2001guidelines}, tend to be rather vague on technical issues, preferring instead to take a ‘rule-of-thumb’ approach. In particular, the guidelines take little or no account of room air flow patterns, which (as discussed above) are crucial to the performance of upper-room UVGI with respect to COVID-19 transmission. Consequently, these guidelines cannot be relied upon in the context of COVID-19 and as a result there is urgent need to develop new guidelines that are focused on this disease.

\subsection{Conclusions}

The conclusions regarding UVGI are:
\begin{itemize}
	\item	There is strong evidence that the SARS-CoV-2 virus can be inactivated by irradiation with UV-C light and that the virus is particularly vulnerable to UV damage when aerosolised.
	\item	There is good theoretical evidence to suggest that upper-room UVGI might be effective at disinfecting the SARS-CoV-2 virus in room air.
	\item	While in theory upper-room UVGI can militate against COVID-19 transmission in enclosed spaces, because larger aerosol particles (i.e. 10-100 $\mu{}$m in diameter) can decouple from room air convection currents, it may be that additional supplementary air movement devices (e.g. room mounted fans, etc.) will be needed to ensure good air disinfection.
	\item	In-duct UVGI air disinfection could potentially be a useful technology for enabling commercial buildings to function during the winter months of the COVID-19 pandemic. 
	\item	There is urgent need for a robust evidence base regarding UVGI air disinfection. In particular, because it is difficult to predict how UVGI equipment will perform in any given context, there is heavy reliance on the claims made by manufacturers. As such, there is urgent need for manufacturers of UVGI equipment to undertake appropriate microbiological testing so that end users can be confident that their products will work effectively against the SARS-CoV-2 virus.
	\item	There is urgent need to develop robust methodologies for validating the performance of UVGI air disinfection systems, in order to demonstrate that they work (i.e. prevent the spread of COVID-19 in buildings).
	\item 	There is a need for suitable guidelines regarding the design and use of UVGI air disinfection systems in the context of COVID-19.
\end{itemize}


\section{Summary of current recommendations for the use of face masks or coverings (August 2020)} \label{app:masks}

Face covering and masks have been demonstrated to prevent the transmission of the COVID-19 virus (\S\ref{sec:masks}) especially in settings where there is reduced physical distancing. Face shields may protect against droplets and the spread of droplets containing the virus, but not the transmission of aerosols. There is emerging evidence that face coverings and masks may provide some protection by reducing the dose of SARS-CoV-2 received. Face coverings have become mandatory within many indoor settings, including shops and on public transport in the UK, where there is risk of decreased social distancing \citep{DHSC2020}. Face masks and coverings may represent the simplest and most cost effective method available to employers to protect their employees if remote/home working is not feasible.

The evidence face masks and coverings will probably improve with time, however public guidance in the UK on masks and face covering usage is likely to change from that written below. Please ensure you check the local and current guidance for up to date information of face coverings. The links / references used below may still be relevant to UK guidance.

\underline{For employers (within offices, etc)}

The UK Government state within their office guidance \citep{BEIS2020} that “there is growing evidence that wearing a face covering in an enclosed space helps protect individuals and those around them from COVID-19”. They also state “minimising time spent in contact, using fixed teams and partnering for close-up work, and increasing hand and surface washing” and that “these other measures remain the best ways of managing risk in the workplace” 

However, given the time spent within other indoor settings, including shops and public transport may be less than within offices, etc., and that distancing is likely to be difficult with an office. Perhaps consideration should be given to the protection of a workforce by the recommending the use face coverings or masks by employers. 

\citep{Jones2020} demonstrates that the amount of virus inhaled by a susceptible person ( \nl{}) in the presence of an infector is particularly affected by the respiratory rate of a susceptible person, the emission rate of RNA copies, the exposure time, the space volume, and the removal rate. A sensitivity analysis shows that predictions of \nl{} in the reference space are most sensitive to the emission rate of \rnac{}. \nl{} is linearly related to the emission rate and so public health messages to encourage self isolation when exhibiting symptoms of COVID-19 and the wearing of face coverings indoors are important.

\underline{Within universities}

No specific advice is given on the UK government for universities. The Scottish Government \citep{ScottishGovernment2020} state: “universities will consider and plan around: availability and public health advice on PPE, other equipment and/or face coverings appropriate to the activity or location.”

Similar to the above situation (for employers) any indoor space represents a risk of transmission. Consideration should be given to the protection of students, teachers, lecturers and other staff by the recommending the use face coverings or masks.

\underline{For schools and colleges}

“Public Health England does not recommend the use of face coverings in nurseries, childminders, schools or colleges”. However, the \citet{ScottishGovernment2020new} in-line with WHO recommendation advises high school children (12 years and older) ``should use face covering between classes, in secondary schools, by adults and all pupils moving around the school, such as in corridors and communal areas where physical distancing is difficult to maintain.'' The guidance \citep{DoE2020} also notes if a child has been using a face covering on public transport or on dedicated school transport that the covering must be removed upon arrival at the school. More recent DofE guidance \citep{DoE2020b} allows discretionary use of face coverings for pupils in Year 7 and above where social distancing cannot be safely managed, and this will be mandated in areas where local restrictions apply.

The risk to children though much lower, is well documented. Their asymptomatic transmission is also well documented, which likely puts the teaching and other staff at increased risk, as well as the parents. Yet it cannot be recommended to go against the advice given. Many European countries have reopened schools without serious issues and without masking children (as of August 2020)

\underline{NOTE:} Advice in Wales for face coverings and masks differs slightly from the other UK countries.

\section{Aerosols in the context of singing, and woodwind \& brass musical instruments}

Music is an important part of our cultural heritage as well as providing or supporting much of the entertainment industry. While recognising the importance of the activities of professional musicians and singers and the organisations which employ them, it is important to note that the student and amateur musical scene represents a much larger number of people.

For professionals, students and amateurs alike, the mechanics of playing or singing are similar, and so measurement of the aerosol risks can be considered in a similar way. There are some significant differences in regard to the impact to the community and environment in which the aerosol is likely to be produced, namely:

\begin{itemize}
\item Professionals and conservatoire level students are likely to be spending greater periods of time playing in both rehearsal and performance settings and be sharing a performance venue with a very much larger and more mixed audience than is the case for amateurs.

\item Outdoor performance by amateur woodwind and brass groups is common, particularly during the summer months. Some amateur groups are choosing to rehearse outdoors at the present time.

\item Shared use or short-term loan of an instrument is common for early level students, and in some amateur groups. This is also an issue for pre-sale trial of instruments.

\item Amateur musical groups include early level students, people of advanced age, and people with pre-existing health conditions.

\end{itemize}

The need for improving the safety of indoor rehearsal and performance is well understood, and research to measure the level of aerosol created during performance is being undertaken, so that risks from musicians and singers can be compared with the risks of other performers such as orators, actors, dancers and sports competitors. Currently, the only published and peer-reviewed report is that on the vuvuzela by \citet{lai2011}. A substantially more thorough study carried out by \citet{reid2020} comparing aerosols produced in singing, speaking and breathing is now available on the pre-print server ChemRxiv. A further piece of work by the same authors investigating aerosol production in musical instruments is expected to be released in the near future \citep{frontrow2020}. Another major study is in preparation at Colorado State University \citep{colorado2020}. More informal studies have been carried out by a number of organisations \citep{bamberger2020, brandt2020, wein2020, kaehler2020,  nfhs2020}.

When such aerosol production is better understood, appropriate ventilation and protection measures for concert halls, theatres and arenas can then be designed. Organisations which represent professionals and amateurs, and also those which represent performance venues, are already producing extensive guidelines on the operational logistics of rehearsal and performance \citep{schutzkonzept2020, bbe2020, mu2020}.

Outdoor performance presents less risk than indoor, particularly where mitigations, such as socially distanced seating of the players and separation from the audience or passers-by, are upheld. Mitigations for indoor playing consider issues of ventilation, and numbers of players in a room of a given volume. Other mitigations that have been suggested, but for which the efficacy has not yet been established, involve the wearing of a mask while playing, with a hole cut for the mouthpiece; and the use of a covering to trap aerosols exiting from the end of the instrument. All such mitigations are undesirable: outdoor performance feels unsatisfactory without the venue reverberation and musicians would usually choose to sit close to each other to hear each other better; wearing a mask while playing will impact on the player's ability to perform; and covers on instruments reduce the sound level and can also impact on tuning and sound quality.

It should also be noted that a person moving through a room can create a wake or cause settled particulates to be resuspended in the air. Musicians who make large body movements may contribute to this effect, so whilst percussion, string players and conductors do not blow their instruments their bodily movements may contribute to the redistribution of aerosol around a rehearsal or performance space. This is illustrated in the string player photographs in \citet{wein2020} where the bow movement entrains the flow.

For woodwind and brass instruments, the player's breath drives the sound production, and breath generated aerosols will be carried through the instrument. Some of that aerosol load would be captured within the instrument as condensation. In brass instruments, which are not perforated with finger holes, and have lengthy and convoluted pipework, such condensation might be expected to be more complete than in shorter woodwind instruments. Notwithstanding this, condensation cannot be relied upon as a means to reduce aerosol when the instrument is warm, or being played in a warm environment.

Since the interior of an instrument can become contaminated if played by an infected player, the issue of effective cleaning of loan or trial instruments is important. An approach might be to place each player on a track and trace register after each loan or trial, or apply a quarantine period to the instrument. If the instrument is needed sooner, or if a player develops COVID-19, then, where feasible, it could be put through a chemical cleaning cycle. Metallic instruments and student grade plastic instruments might tolerate such treatment, but quality wooden instruments would not.

There is considerable ongoing research in this area at the present time, and it may be some time before complete peer-reviewed reports are available. Some reports cited here are media press reports.


\clearpage
\bibliographystyle{jfmF}
\bibliography{reference}

\end{document}